\documentclass[twocolumn]{raa}
 
\usepackage{graphicx,times}             
\usepackage{natbib}
\usepackage{amssymb,amsmath}
\usepackage{mathrsfs}
\bibpunct{(}{)}{;}{a}{}{,}

\usepackage[pagebackref=true]{hyperref}

\begin{document}

\title{New simulations of the X-ray spectra and polarizations of accretion-disc corona systems with various geometrical configurations I. Model Description}

   \volnopage{Vol.0 (20xx) No.0, 000--000}     
   \setcounter{page}{1}           

   \author{Xiao-Lin Yang  
      \inst{1,2,3,4}
   \and Jian-Cheng Wang
      \inst{1,2,3,4}
   \and Chu-Yuan Yang 
      \inst{1,2,3}
   }

   \institute{Yunnan Observatories, Chinese Academy of Sciences,
396 Yangfangwang, Guandu District, Kunming, 650216, P. R. China; {\it yangxl@ynao.ac.cn}\\ 
        \and
             Key Laboratory for the Structure and Evolution of
Celestial Objects, Chinese Academy of Sciences, 396 Yangfangwang,
Guandu District, Kunming, 650216, P. R. China; {\it jcwang@ynao.ac.cn}\\
        \and
             Center for Astronomical Mega-Science, Chinese Academy of Sciences,
20A Datun Road, Chaoyang District, Beijing, 100012, P. R. China;\\
        \and 
             University of Chinese Academy of Sciences, Beijing, 100049, P. R. China\\
\vs\no
   {\small Received 20xx month day; accepted 20xx month day}}

\abstract{Energetic X-ray radiations emitted from various accretion systems are widely considered to be produced by 
Comptonization in the hot corona. The corona and its interaction with the disc play an essential role in the evolution of the system and are potentially responsible for many observed features. However many intrinsic properties of the corona are still poorly understood, especially for the geometrical configurations. The traditional spectral fitting method is not powerful enough to distinguish various configurations. In this paper we intent to investigate the possible configurations by modeling the polarization properties of X-ray radiations. The geometries of the corona include the slab, sphere and cylinder. The simulations are implemented through the publicly available code, LEMON, which can deal with the polarized radiative transfer and different electron  distributions readily. The results demonstrate clearly that the observed polarizations are dependent on the geometry of the corona heavily. The slab-like corona produces the highest polarization degrees, the following are the cylinder and sphere. One of the interesting things is that the polarization degrees first increase gradually and then decrease with the increase of photon energy.  For slab geometry there exists a zero point where the polarization vanishes and the polarization angle rotates for $90^\circ$. These results may potentially be verified by the upcoming missions for polarized X-ray observations, such as $IXPE$ and $eXTP$.
}

   \authorrunning{X.-L. Yang, J.-C. Wang \& C. Y. Yang }             
   \titlerunning{Modeling of X-ray spectra and polarizations}  

   \maketitle

\section{Introduction}
\label{Sec:A: Introduction} 
Active galactic nuclei (AGNs), $\gamma$-ray bursts (GRBs) and X-ray binaries are the most powerful X-ray objects in the universe. Accreting ambient materials and then releasing gravitational energy by the central compact objects is one of the most energetic phenomena in astrophysics. The released energies will eventually heat the accreting gases and result in the radiations range from radio to gamma-rays. The X-rays are widely believed to be produced by the inverse Comptonization of soft photons in the corona which is a hot region close to the central objects (e.g. \cite{haardt1991, haardt1993, 1975ApJ...199L.153E, 1975ApJ...195L.101T, 2010LNP...794...17G}). And the soft photons are usually multi-temperature blackbody emissions and come from the accretion disc (\cite{1973A....24..337S,1974ApJ...191..499P,1988ApJ...332..646A,2014ARA..52..529Y}).

The corona plays a very important role in the disc-corona system. While due to the complicated physical processes involved in the accretion system, the evolution, formation and heating of the corona and its geometrical configurations are still under debate (e.g. see the discussions given by \cite{2021ApJ...906...18D,2022MNRAS.510.3674U,2021NatCo..12.1025Y}). Various physical processes can lead to the formation of a corona and they show similar spectral profiles or spectral energy distribution (SED). For example the corona with an extended slab-like geometry is usually sited above the disc and may be a result of magnetic instabilities \citep{1979ApJ...229..318G,1998MNRAS.299L..15D}. Materials accreted around a neutron star instead of a black hole will accumulate and finally form into a transition layer which shows the characteristics of a corona \citep{2000A...358..617S, 2022ApJ...924L..13L}. At the vicinity of the black hole, the quite active magnetic processes could release considerable energies by magnetic reconnection which can heat the plasma into a very high temperature (e.g. \cite{2012MNRAS.424.1284W}). The corona can even be formed by the evaporation of the inner part of an accretion disc, or as the transfer region between the jet and the black hole, either as a failed jet \citep{2004A...413..535G} or as a standing shock wave \citep{1991ApJ...374..741M, 1999ApJ...519L.165F, 2007ARv..15....1D}. The geometrical configurations of these corona models are deeply connected with their physical origins. Thus, to distinguish the geometry of corona from observable quantities will provide significant constrains on the physics of the accretion system (e.g. \cite{2022MNRAS.510.3674U, 2022ApJ...924L..13L, 2021ApJ...906...18D}).

Former researches have put some constrains on the geometry of corona. For example, the size and location of X-ray corona have been estimated to be within a few gravitational radii by the microlensing observations \citep{2004ApJ...605...58K, 2013ApJ...769L...7R, 2016AN....337..356C}. Comparing the time lags between the direct and reflected radiations (or radiations in different energy bands) can provide further constrains on the geometrical parameters of the disc-corona system (e.g. \cite{2019MNRAS.488..324I, 2021MNRAS.507...55M}). However, the polarization of X-ray radiations is an alternative and unique way to give possible new constrains on the corona geometry \citep{2021ApJ...906...18D, 2022ApJ...924L..13L, 2022MNRAS.510.3674U, 2021NatCo..12.1025Y}, since the polarizations induced by the inverse Comptonization are intrinsically dependent on geometry and electron distribution  \citep{2010ApJ...712..908S, 2011Sci...332..438L, 2017ApJ...850...14B}. The Compton scattering can be simply divided as Thomson and Klein-Nishina regimes according to the energy of incident photons and the cross sections are intrinsically polarization dependent \citep{Fano1949, 1960ratr.bookC}. It can induce polarizations for anisotropic and unpolarized photons that scatter off non-relativistic electrons \citep{1970A.....7..292B, 2009ApJ...701.1175S}. For photons scattering off energetic electrons the polarization will be suppressed due to the beaming effect (e.g. \cite{2021ApJ...906...18D}). Thus the polarized radiations among the X-ray bands would be reasonably expected \citep{1993A...275..337P, 2010ApJ...712..908S, 2017ApJ...850...14B, 2022MNRAS.510.3674U}  and they can be used as a useful probe to distinguish the geometrical configurations of the corona-disc systems.

Following the previous studies, we are here motivated to provide constrains on the corona geometry by modeling the observed X-ray spectra and polarizations. Our paper is organized as follows. The model and method are introduced in Section \ref{Sec: ModelandMethod}. In the Section \ref{Sec: results}, we present the results of our calculations. The discussions and conclusions are finally provided in Section \ref{Sec: discussions}.

\section{Model and Method}
\label{Sec: ModelandMethod}

In this section, we give an introduction to the model and method used in this paper, which are based on our public available code Lemon \citep{2021ApJS..254...29X}. Here we mainly discuss how to generate photons effectively in these geometrical configurations for the scattering, and show the estimation procedures.

\subsection{the Geometries of the Corona}

\begin{figure}[t!]
\center 
\includegraphics[scale = 1.0]{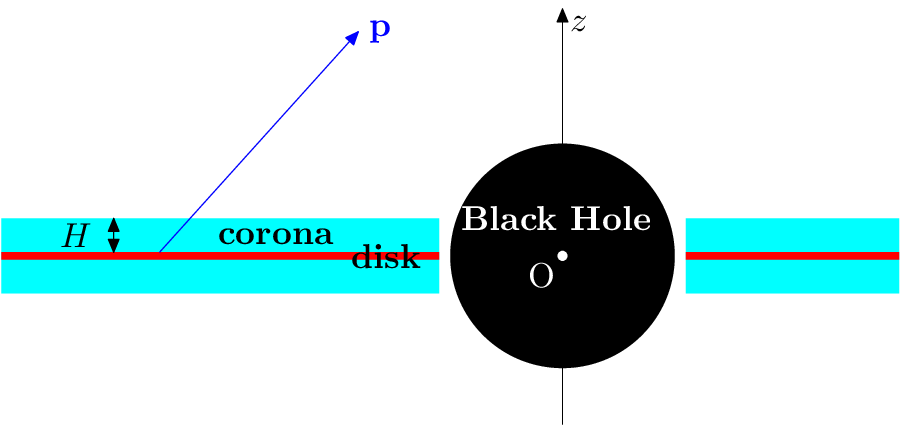}  
\includegraphics[scale = 1.0]{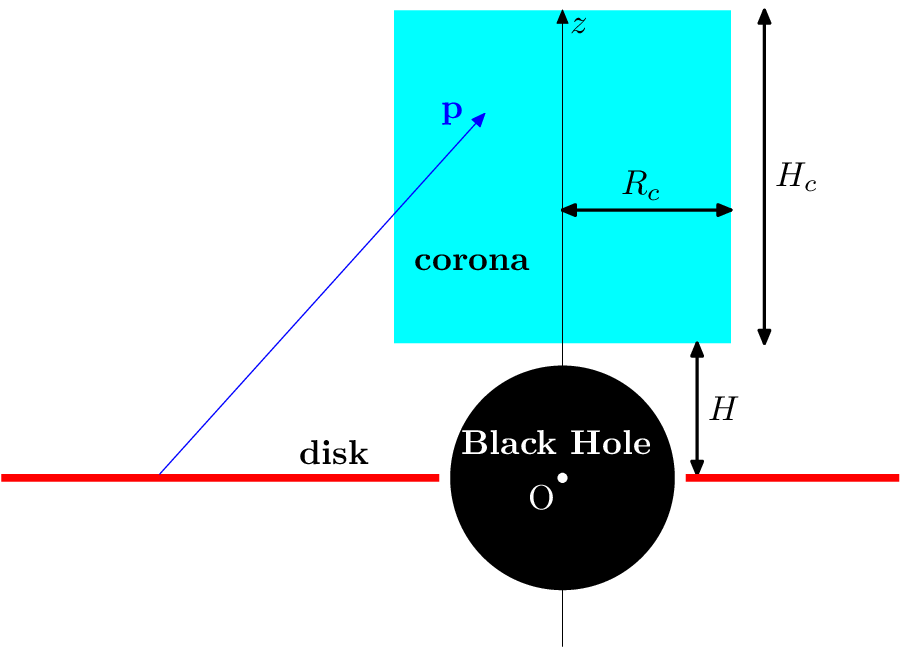}  
\caption{The schematic diagrams for the geometries of the corona with a slab (top panel) and cylinder (bottom panel) configurations. The red and blue regions represent the accretion disc and the corona, respectively. The black hole is represented by the central black disc.}
\label{fig1}
\end{figure}

\begin{figure}
\center 
\includegraphics[scale = 0.6]{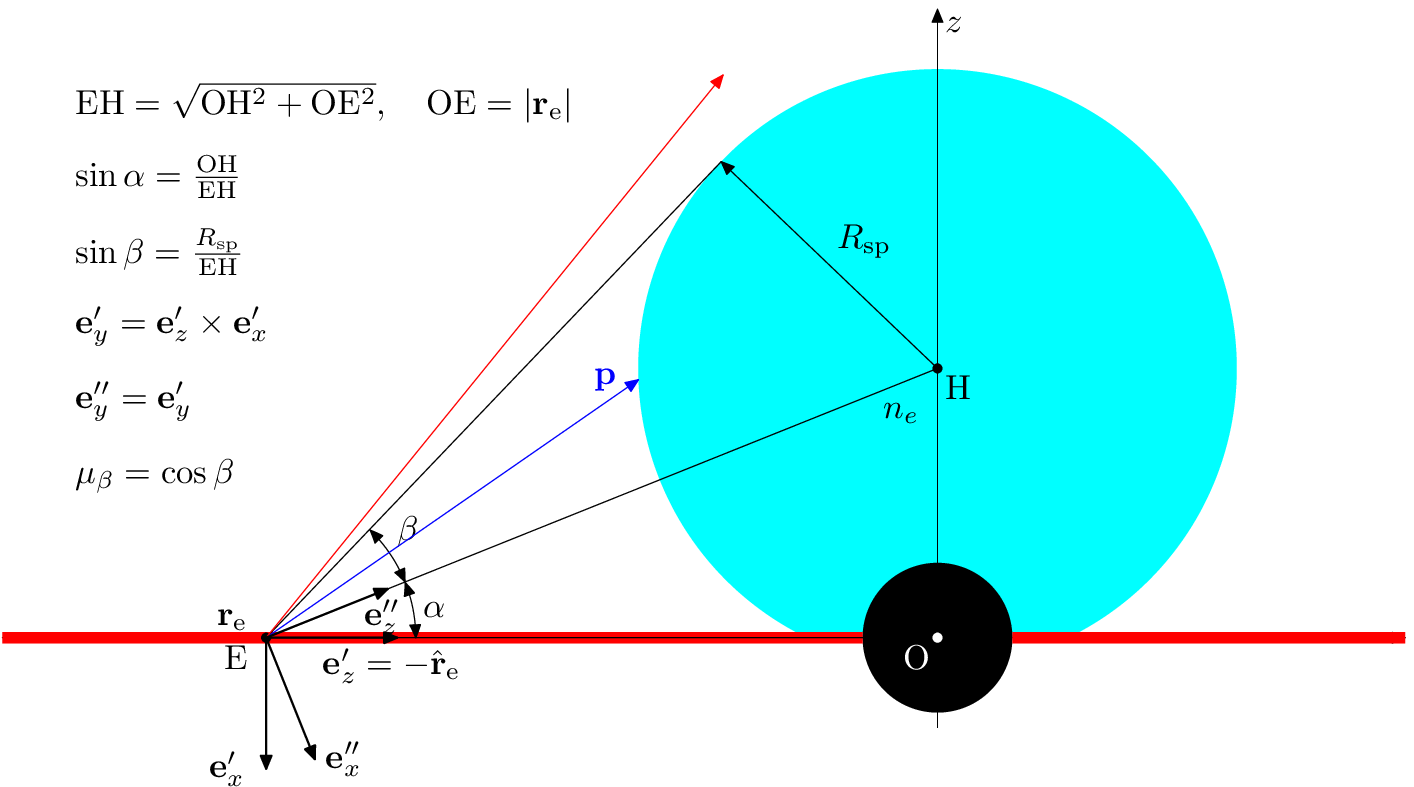} 
\includegraphics[scale = 0.6]{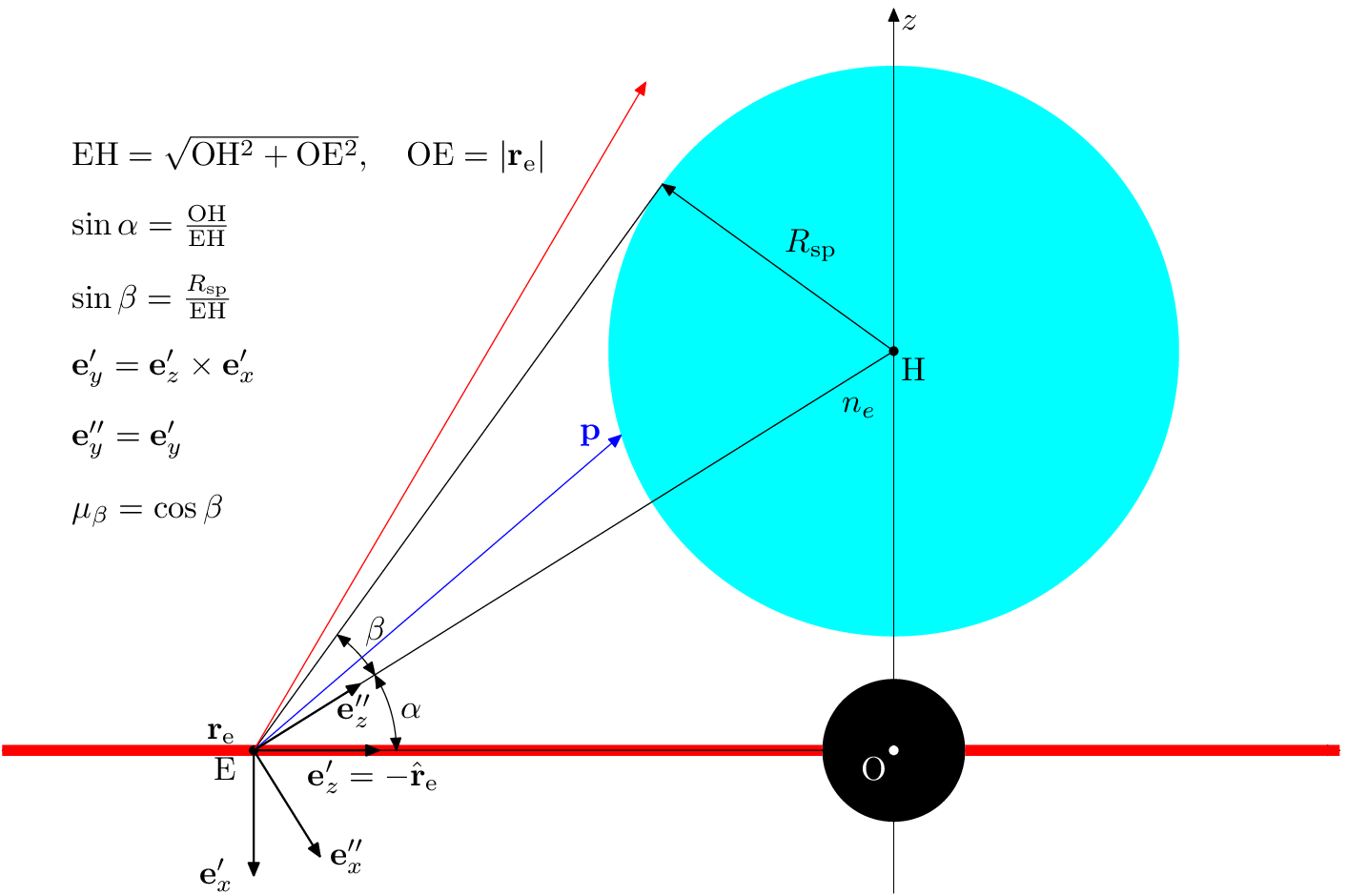} 
\includegraphics[scale = 0.6]{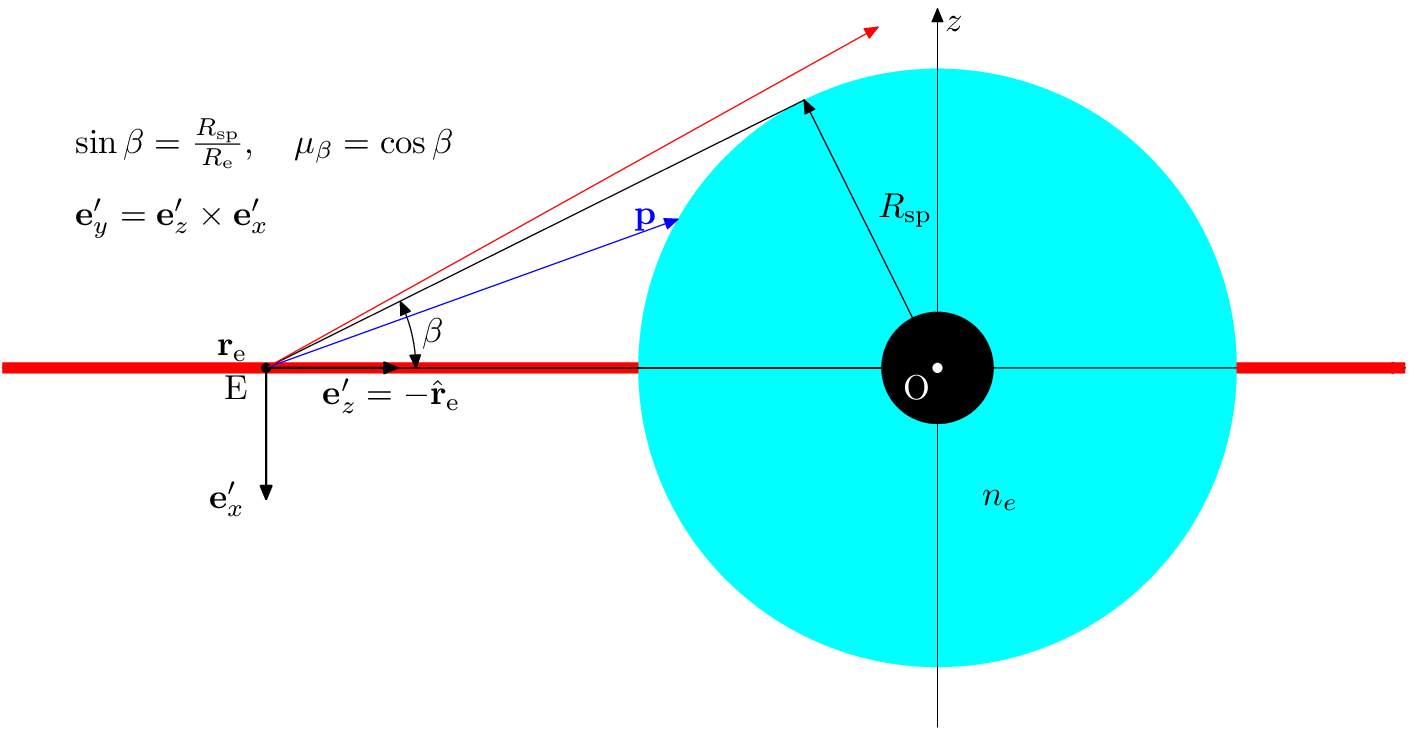}   
\caption{The schematic descriptions for the geometrical relationships between the spherical corona and the accretion disc for various heights $H$ of sphere center. The red and blue regions represent the accretion disc and the corona, respectively. Seed photons are emitted from the disc surface and only those with their momentum direction $\mathbf{p}$ (the blue vector) falling into the cone of $\beta$ can reach the corona atmosphere. Paths of photons with a red momentum vector will not intersect with the corona and should be rejected.}
\label{fig2}
\end{figure}

In this paper, we will calculate the observed spectra and polarizations from three kinds of coronas with slab, sphere and cylinder geometries,  respectively. Their geometrical configurations are shown in Figure \ref{fig1} and \ref{fig2}. The slab-like corona is a thin layer and sandwiches the accretion disc completely. Its geomertry is determined by the inner and outer radii: $R_{\rm in},\,\,R_{\rm out}$, and the height $H$. The cylinder-like corona is located above the disc with the top and bottom surfaces putting at height $H$ and $H+H_c$. The radius of the cylinder is $R_c$. As the top surface of the cylinder is set to be sufficiently high, $H_c$ is also very large, thus the effects of its minor changes on the results can be ignored. The corona with a sphere geometry is shown in Fig. \ref{fig2}, which is described by the sphere radius $R_{\rm sp}$ and the height $H$ of the sphere center with respect to the disc. 

The physical parameters to describe these coronas are the electron temperature $kT_e$, the number density $n_e$ and the optical depth $\tau$ of Thomson scattering. For the sake of simplicity, both $kT_e$ and $n_e$ for all of three kinds of coronas are set to be constants throughout the corona. The values of the Thomson optical depth $\tau$ for three cases are fixed and given by:
$\tau = \sigma_T n_e H, \tau = \sigma_T n_e R_c, \tau = \sigma_T n_e R_{\rm sp}$, respectively, where $\sigma_T$ is the Thomson cross section. Hence, as $\tau$ and $H, R_c, R_{\rm sp}$ are provided, one can calculate the electron number density by
\begin{equation}
    n_e = \frac{\tau}{\sigma_T H}, \,\,n_e = \frac{\tau}{\sigma_T R_c}, \,\,n_e = \frac{\tau}{\sigma_T R_{\rm sp}},
\end{equation}
and vice versa.

\subsection{Photon Generation}

For all the configurations, the low energy seed photons are emitted from the standard geometrically thin and optically thick accretion disc \citep{1973A....24..337S}. We assume that the accretion disc is in a multi-temperature state and its surface temperature changes with the disc radius $R$. The distribution function of the temperature is given by \citep{1973A....24..337S, 2018A...619A.105T}
\begin{eqnarray}
\begin{split}
   T(R) = \left[\frac{3GM\dot{m}}{8\pi R^3 r_g^3 \sigma_{\rm SB}}\left(1-\sqrt{\frac{R_{\rm in}}{R}}\right)\right]^{1/4},
\end{split}
\end{eqnarray}
where $r_g$ is the gravitational radius, $M$ is the mass of the central black hole, $G$ is the gravitational constant, $\sigma_{\rm SB}$ is the Stefan-Boltzmann constant and $\dot{m}$ is the accretion rate. The radius $R$ is in unit of $r_g$. For convenience, we will use the energy release rate $\eta$ and the Eddington luminosity $L_{\rm EDD}$ to replace $\dot{m}$ as $\eta L_{\rm EDD} = \dot{m}c^2$, where $c$ is the light speed. The Eddington luminosity is defined by  $L_{\rm EDD}=4\pi GM m_pc/\sigma_T$, where $m_p$ is the proton mass \citep{2014ARA..52..529Y}. Then the expression of $T(R)$ can be rewritten as
\begin{eqnarray}
\begin{split}
   T(R) = \left[\frac{ \eta C}{ R^3 }\left(1-\sqrt{\frac{R_{\rm in}}{R}}\right)\right]^{1/4},
\end{split}
\end{eqnarray}  
where $C=3 L_{\rm EDD}/(8\pi r_g^2 \sigma_{\rm SB})$. In all the simulations, the values of $\eta$ and $M$ are given in advance and set to be $0.1$ and 10 $M_{\sun}$, corresponding to a mass accretion rate of $\dot{m}\approx 1.40 \times 10^{18}$ g$/$s (see table \ref{table1}).

In order to describe the Keplerian motions of the accretion disc, we define a static reference frame, whose basis vectors are given by $\mathbf{e}_x$, $\mathbf{e}_y$ and $\mathbf{e}_z$. With $\mathbf{e}_i$, we can further define a local reference frame at radius $R$ and azimuth angle $\varphi$, the basis vectors of which are constructed by
\begin{eqnarray}\label{triad_slab1}
\left\{\begin{array}{rl}
  \mathbf{e}'_x &= - (\cos\varphi\mathbf{e}_x + \sin\varphi\mathbf{e}_y),\\
  \mathbf{e}'_y &= \sin\varphi\mathbf{e}_x - \cos\varphi\mathbf{e}_y , \\
  \mathbf{e}'_z &= \mathbf{e}_z.
\end{array}\right.
\end{eqnarray}  
Then with respect to the static reference frame, the Keplerian velocity of the disc at $R$ and $\varphi$ can be expressed as
\begin{eqnarray}\label{eqa:v_kepler}
\begin{split}
   \mathbf{V}_k(R, \varphi) = -\frac{c}{\sqrt{R}}\mathbf{e}'_y.
\end{split}
\end{eqnarray}  

Within the comoving frame of the disc at $(R, \varphi)$, we assume that the emissivity $\widetilde{j}$ along direction $\widetilde{\mathbf{\Omega}}=(\widetilde{\theta}, \widetilde{\phi})$ and at frequency $\widetilde{\nu}$ is a modified black body radiation given by 
\begin{eqnarray}
\begin{split}
   \widetilde{j}(R, \varphi, \widetilde{\mathbf{\Omega}}, \widetilde{\nu}) = \frac{2h\widetilde{\nu}^3}{c^2}\frac{1+a\widetilde{\mu}}{e^{h\widetilde{\nu}/k_{\rm B}T(R)}-1},
\end{split}
\end{eqnarray}
where $a=2.06$ is a constant and $\widetilde{\mu}=\cos\widetilde{\theta}$. Factor $1+a\widetilde{\mu}$ means that the emissivity obeys the limb-darkening law \citep{2018A...619A.105T}. Then the corresponding emissivity $j$ in the static reference can be obtained through a Lorentz transformation \citep{1973erh..book.....P}
\begin{eqnarray}\label{eqa_emissivity}
\begin{split}
   j(R, \varphi, \mathbf{\Omega}, \nu) = (\gamma D_p)^2\widetilde{j}(R, \varphi, \widetilde{\mathbf{\Omega}}, \widetilde{\nu}),
\end{split}
\end{eqnarray}
where $\gamma=1/\sqrt{1-V_k^2/c^2}$ is the Lorentz factor, $D_p=1+\widetilde{\mathbf{\Omega}}\cdot\mathbf{V}_k/c$ and 
\begin{eqnarray}
\begin{split}
    \widetilde{\nu} &= \nu\gamma\left(1-\frac{\mathbf{\Omega}\cdot\mathbf{V}_k}{c}\right),\\
    \widetilde{\mathbf{\Omega}} &= \frac{1}{\gamma D_m}\left[\mathbf{\Omega}-\frac{\gamma(1+\gamma D_m)}{1+\gamma}\frac{\mathbf{V}_k}{c}\right],
\end{split}
\end{eqnarray}
where $D_m=1-\mathbf{\Omega}\cdot\mathbf{V}_k/c$. Notice that the quantities related to the Lorentz transformation are defined with respect to the frame $\mathbf{e}'_i$ given by Equation \eqref{triad_slab1}, using Equation \eqref{eqa:v_kepler}, the above equations can be written explicitly as
\begin{eqnarray}
\left\{\begin{array}{rl}
    \widetilde{\nu} &= \displaystyle\nu\gamma\left(1+\Omega_y\beta\right),\\
    \widetilde{\Omega}_x &= \displaystyle\frac{\Omega_x}{\gamma D_m},\\ 
    \widetilde{\Omega}_y &= \displaystyle\frac{1}{D_m}(\Omega_y+\beta),\\
    \widetilde{\Omega}_z &= \displaystyle\frac{\Omega_z}{\gamma D_m}, 
\end{array}\right.
\end{eqnarray}
where $\beta=V_k/c=1/\sqrt{R}$, $D_m=1+\Omega_y\beta$ and $D_p=1-\widetilde{\Omega}_y\beta$.

In our former paper \citep{2021ApJS..254...29X} we have explained that the Monte Carlo radiative transfer is actually equivalent to the evaluation of the Neumann solution of the radiative transfer equation. Each term of Neumann solution is a multiple integral which can be written as
\begin{equation} 
\begin{split}
    I_m=\int n(P_0)K(P_0\rightarrow P_1)\cdots K(P_{m-1}\rightarrow P_m) f(P_m) dP_0 dP_1\cdots dP_m,
\end{split}
\end{equation} 
where $n(P_0)$ is the emitted photon number density and related to the emissivity $j(P_0)$ through $n(P_0)=j(P_0)/h\nu$, where $h$ is the Plank constant, $K(P\rightarrow P')$ is the transfer kernel, $f(P_m)$ is the recording function and $P=(\mathbf{r}, \mathbf{\Omega}, \nu)$, which are the position and momentum vectors and frequency of the photon, respectively. The generation of photons are actually related to the calculation of the integral in terms of $P_0$ by Monte Carlo method, which can be separately written as 
\begin{equation} \label{eqa_integral}
\begin{split}
    &\int n(\mathbf{r}, \mathbf{\Omega}, \nu)\delta(z) d\mathbf{r}d\mathbf{\Omega}d\nu,
\end{split}
\end{equation}
where a $\delta$-function $\delta(z)$ is inserted, since the seed photons are emitted by the disc that is located on the equatorial plane. We assume that $\nu_1\le\nu\le\nu_2$, since for sufficient small and large frequency $\nu$, the contributions from the black body radiation can be ignored. $\nu_1$ and $\nu_2$ are free parameters and set to be $h\nu_1 = 10^{-6} m_ec^2$, $h\nu_2 = 10^{-2} m_ec^2$, respectively. For convenience, we introduce a new variable $y$ to replace $\nu$ and $\nu=10^y$, $y_1\le y\le y_2$, where $y_1=\log_{10}(\nu_1)$, $y_2=\log_{10}(\nu_2)$. Then Equation \eqref{eqa_integral} can be rewritten as
\begin{equation}\label{integralJ}
\begin{split}
    &\int n(\mathbf{r}, \mathbf{\Omega}, \nu)\delta(z) d\mathbf{r}d\mathbf{\Omega}d\nu,\\
   =&\ln10\int n(R, \varphi; \mu, \phi; \nu)\nu R dR d\varphi d\mu d\phi dy.
\end{split}
\end{equation}

To utilize Monte Carlo method to evaluate the above integral, we split the integrand into two parts, one is used as PDFs for $R, \varphi, \mu, \phi$ and $y$, the other one is used as weight for the integral. For $R$, $\varphi$ and $y$, we assign them with the PDFs given by
\begin{equation} \label{eqa3} 
\left\{\begin{array}{lr}
    p(R)=\displaystyle \frac{R}{\mathcal{N}_R},& \,\, R_{\rm in}\le R\le R_{\rm out},\\
    p(\varphi)=\displaystyle \frac{1}{2\pi},& \,\, 0\le \phi\le 2\pi,\\
    p(y)=\displaystyle \frac{1}{y_2-y_1},& \,\, y_1\le y\le y_2,
\end{array}\right. 
\end{equation}
where $\mathcal{N}_R=(R^2_{\rm out}-R_{\rm in}^2)/2$ is the normalization factor. Then $R, \varphi$ and $y$ can be sampled directly by 
\begin{equation} 
   R_e = \sqrt{R_{\rm in}^2 + 2\mathcal{N}_R\xi_1},\quad 
   \varphi_e = 2\pi\xi_2,\quad y_e=y_1+(y_2-y_1)\xi_3, 
\end{equation}
where $\xi_i$ are random numbers, whose PDFs are $p(\xi)=1$ and $0\le\xi\le1$. From now on, we will use $\xi$ to represent random numbers, unless otherwise stated. With $R_e, \varphi_e$ and $y_e$, the position vector of the emission site and the frequency of the photon can be obtained as
\begin{eqnarray}
\begin{split}
   \mathbf{r}_e &= R_e(\cos\varphi_e\mathbf{e}_x + \sin\varphi_e\mathbf{e}_y),\\
   \nu_e &= 10^{y_e}.
\end{split}
\end{eqnarray}

The sampling of the initial direction $\mathbf{\Omega}_e=(\mu_e, \phi_e)$ is more complicated and will be discussed in the following sections for the three geometrical configurations, respectively.

While in order to discuss the initial weight $w_{\rm ini}$, we suppose that $(\mu_e, \phi_e)$ has already been obtained. Then $w_{\rm ini}$ equals the remaining part of the integrand of Equation \eqref{integralJ}, i.e., 
\begin{eqnarray}
\begin{split}
   w_{\rm ini} &= \ln10\frac{n(R_e, \varphi_e; \mu_e, \phi_e; \nu_e)\nu_e R_e}{p(R_e)p(\varphi_e)p(y_e)},\\
   & = \mathcal{C}j(R_e, \varphi_e; \mu_e, \phi_e; \nu_e),
\end{split}
\end{eqnarray}
where $\mathcal{C}= \mathcal{N}_R 2\pi (y_2-y_1)\ln10/h$ and $j$ is given by Equaton \eqref{eqa_emissivity}.

\subsubsection{Slab Case}

For the slab corona, the procedure is quite simple, since any photons emitted by the disc will enter the corona automatically.  We first construct a local triad at $\mathbf{r}_{\rm e}$ by
\begin{eqnarray}\label{triad_slab}
\begin{split} 
  \mathbf{e}'_x = - \hat{\mathbf{r}}_{\rm e}, \quad\mathbf{e}'_z = \mathbf{e}_z, \quad 
  \mathbf{e}'_y = \mathbf{e}'_z \times \mathbf{e}'_x,
\end{split}
\end{eqnarray} 
where $\hat{\mathbf{r}}_e=\mathbf{r}_e/R_e$ is the unit vector of $\mathbf{r}_{\rm e}$. With this triad one can obtain $\mathbf{\Omega}_e=(\mu_e, \phi_e)$ directly
\begin{eqnarray} 
\begin{split}
    \mu_e=\xi_1, \quad \phi_e=2\pi\xi_2,
\end{split}
\end{eqnarray} 
and
\begin{eqnarray}\label{pvec}
\begin{split}
   \mathbf{\Omega}_e= \sqrt{1-\mu^2_e}\cos\phi_e \mathbf{e}'_x + \sqrt{1-\mu^2_e}\sin\phi_e \mathbf{e}'_y + \mu_e \mathbf{e}'_z.
\end{split}
\end{eqnarray}

\subsubsection{Cylinder Case}

\begin{figure}
\center  
\includegraphics[scale = 0.9]{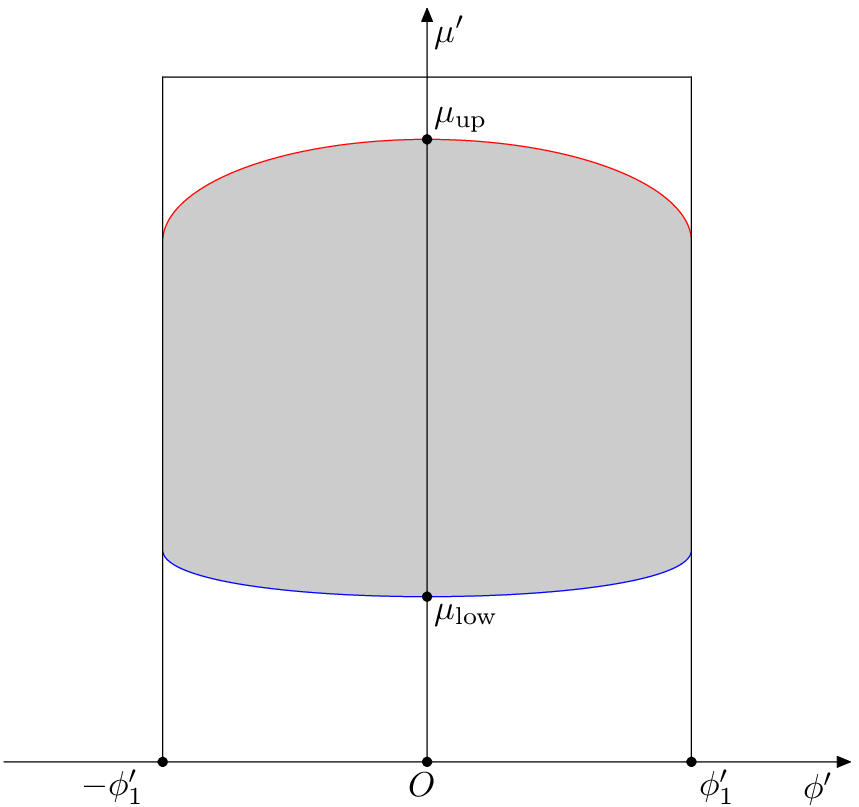}  
\includegraphics[scale = 0.8]{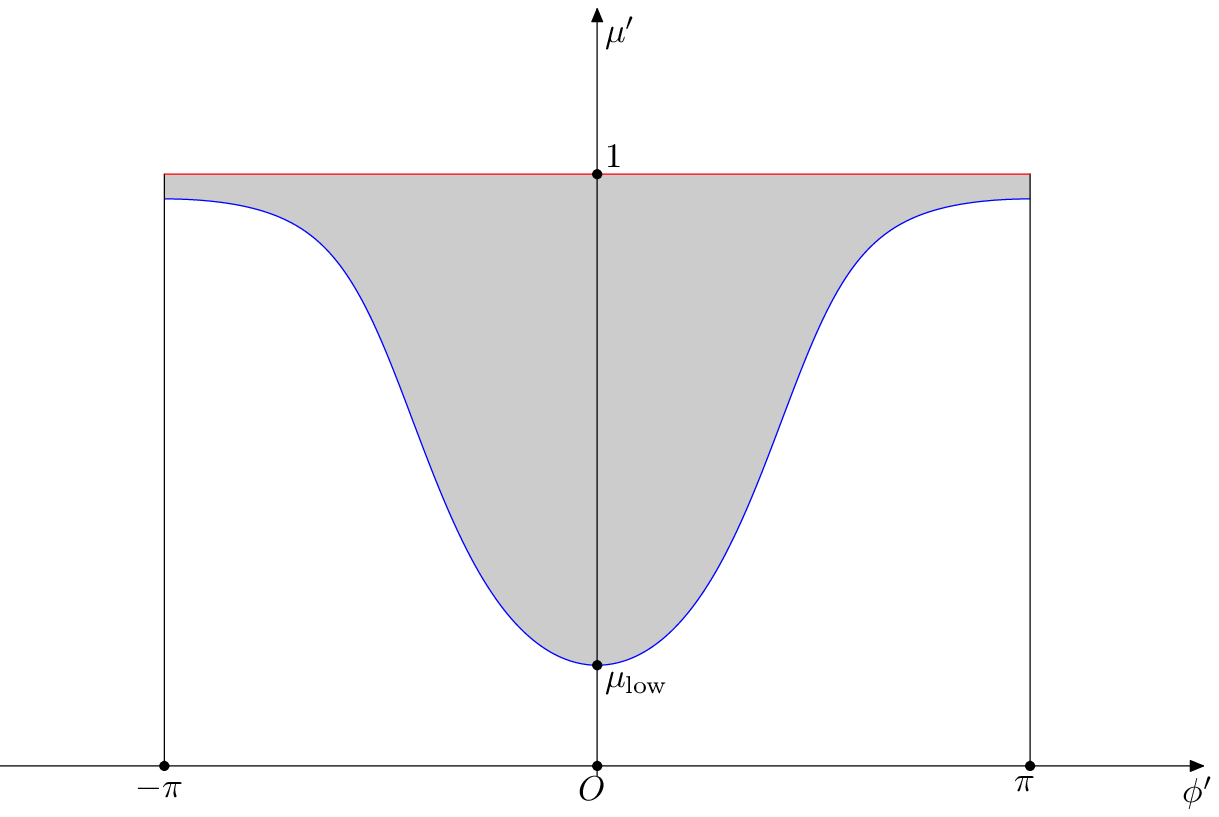}  
\caption{The region formed by the effective photon momentum directions in the $\phi'$-$\mu'$ plane of the triad defined by Equation \ref{triad_slab} for the cylinder case. The top panel shows the case as $R_e>R_c$ and the bottom panel as $R_e\le R_c$. The functions of the boundary curves of this region are given by Equation \ref{mufunc}. A sample of momentum directions will be accepted if it falls into the grey region, otherwise it will be rejected. The sampling algorithms are given by Equations \ref{sample_cylinder} and \ref{sample_cylinder2}. The area $S(R_e)$ of this region is a function of $R_e$ and should be multiplied with $w_{\rm ini}$ to regulate the weight.}
\label{fig4}
\end{figure}
 
Comparing to the slab, the sampling procedures of $\mathbf{\Omega}_e$ for the cylinder and sphere cases are more complicated. This is because if we sample the emission direction $\mathbf{\Omega}_e$ isotropically in $\mathbf{e}'_{xyz}$, the efficiency will be quite low, since many photon samples will miss the cylinder (or sphere) directly. To increase the efficiency, for the cylinder case, we need to get the region formed by the effective directions of the photon in the $\mu'_e$-$\phi'_e$ plane. Here, a direction denoted by $\mathbf{\Omega}_e$ is effective, it means that a photon assigned with this direction can reach the corona eventually. Using the geometrical definitions given in Figure \ref{fig1}, we can obtain the region in the $\mu'_e$-$\phi'_e$ plane directly, which is shown in Figure \ref{fig4}. As $R_e>R_c$ (top panel of Figure \ref{fig4}) the blue and red curves are the boundaries of this region and their function expressions are given by
\begin{eqnarray}\label{mufunc}
\begin{split}
   \mu'_{\rm low}(\phi') & = \frac{H}{\sqrt{\left(R_e\cos\phi'+
        \sqrt{R_c^2-R_e^2\sin^2\phi'}\right)^2+H^2}}, \,\, -\phi'_1\le \phi'\le\phi'_1,\\
   \mu'_{\rm up}(\phi') & = \frac{H+H_c}{\sqrt{\left(R_e\cos\phi'-
        \sqrt{R_c^2-R_e^2\sin^2\phi'}\right)^2+(H+H_c)^2}}, \,\, -\phi_1\le \phi'\le\phi_1,
\end{split}
\end{eqnarray}
where $\phi'_1=\arccos(\sqrt{R_e^2-R_c^2}/R_e)$. Then the direction is sampled by
\begin{eqnarray}\label{sample_cylinder}
\begin{split}
   &{\rm do} \\
   &\quad\quad\mu' = \mu_{\rm low} + (\mu_{\rm up} - \mu_{\rm low})\xi_1, \quad
   \phi' = -\phi'_1 + 2\phi'_1 \xi_2,\\
   &\quad\quad{\rm if( \mu'_{\rm low}(\phi') \le \mu' \,\,and \,\,\mu'\le \mu'_{\rm up}(\phi') )exit},\\ 
   &{\rm end\,\, do} 
\end{split}
\end{eqnarray}
where 
\begin{eqnarray} 
\begin{split}
   \mu_{\rm low}=\mu'_{\rm low}(0) & = \frac{H}{\sqrt{\left(R_e + R_c\right)^2+H^2}}, \\
   \mu_{\rm up}=\mu'_{\rm up}(0) & = \frac{H+H_c}{\sqrt{\left(R_e - R_c\right)^2+(H+H_c)^2}},
\end{split}
\end{eqnarray}
which are the minimum and maximum of two functions given by Equation \ref{mufunc}. As $R_e\le R_c$ (bottom panel of Figure \ref{fig4}), the functions of boundary curves become
\begin{eqnarray}\label{mufunc}
\begin{split}
   \mu'_{\rm low}(\phi')&= \frac{H}{\sqrt{\left(R_e\cos\phi'+
        \sqrt{R_c^2-R_e^2\sin^2\phi'}\right)^2+H^2}}, \,\, &-\pi\le \phi'\le\pi,\\
   \mu'_{\rm up}(\phi')&=1 , \,\, &-\pi\le \phi'\le\pi,
\end{split}
\end{eqnarray}
and the sampling procedure becomes
\begin{eqnarray}\label{sample_cylinder2}
\begin{split}
   &{\rm do} \\
   &\quad\quad\mu' = \mu_{\rm low} + (1 - \mu_{\rm low})\xi_1, \quad
   \phi' = -\pi + 2\pi \xi_2,\\
   &\quad\quad{\rm if( \mu'_{\rm low}(\phi') \le \mu' )exit},\\ 
   &{\rm end\,\, do} 
\end{split}
\end{eqnarray}

Once $\mu'$ and $\phi'$ are obtained, $\mathbf{\Omega}_e$ can be constructed from Equations \eqref{pvec}.

There is a subtle thing that needs to be clarified, i.e., to make the final results correct one must update the initial weight $w_{\rm ini}$ by multiplying a factor $S(R_e)$, namely, $w'_{\rm ini}=w_{\rm ini}S(R_e)$. And $S(R_e)$ is the area of the grey region in Figure \ref{fig4} and can be calculated by
\begin{eqnarray}\label{deltaSmu1}
   \begin{split} 
      S(R_e) =
        \left\{\begin{array}{lll}
           &\displaystyle \int_{-\phi'_1}^{\phi'_1}\left[\mu'_{\rm up}(\phi')-\mu'_{\rm low}(\phi')\right]d\phi', 
                         \quad & R_e > R_c,\\ 
           &\displaystyle 2\pi - \int_{-\pi}^{\pi} \mu'_{\rm low}(\phi') d\phi', \quad & R_e\le R_c.
        \end{array}\right.
   \end{split}
\end{eqnarray}

\subsubsection{Sphere Case}

\begin{figure}
\center 
\includegraphics[scale = 0.65]{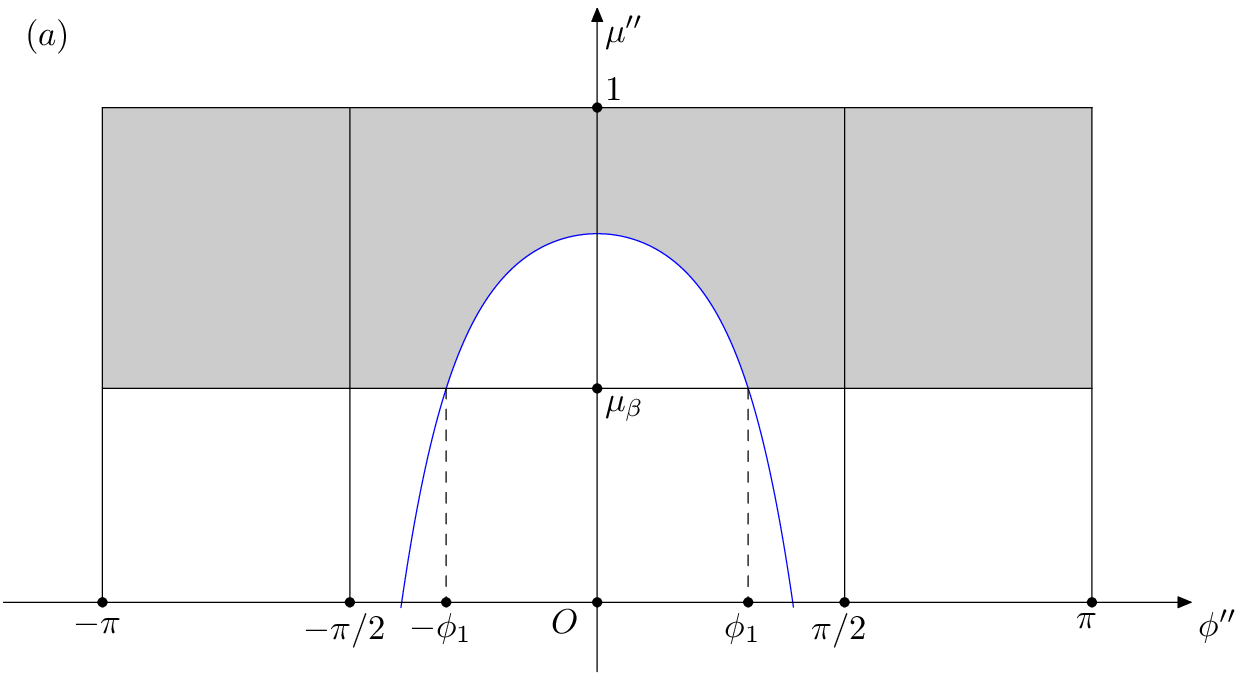} 
\includegraphics[scale = 0.65]{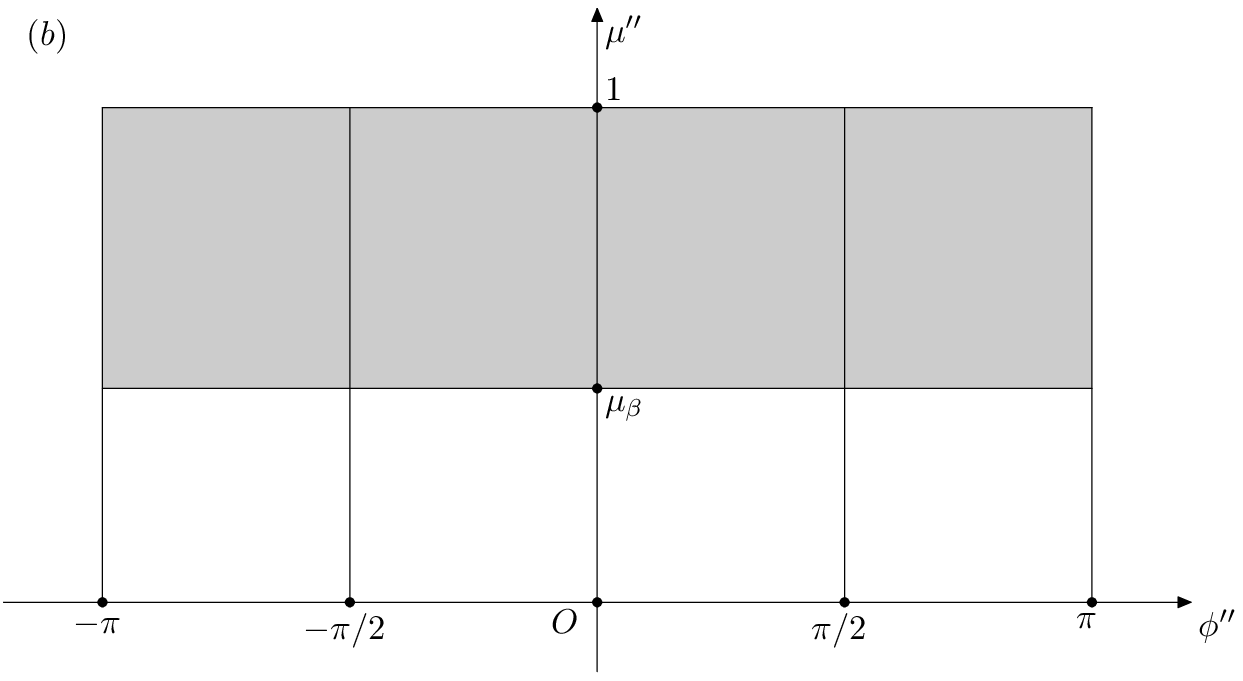} 
\includegraphics[scale = 0.65]{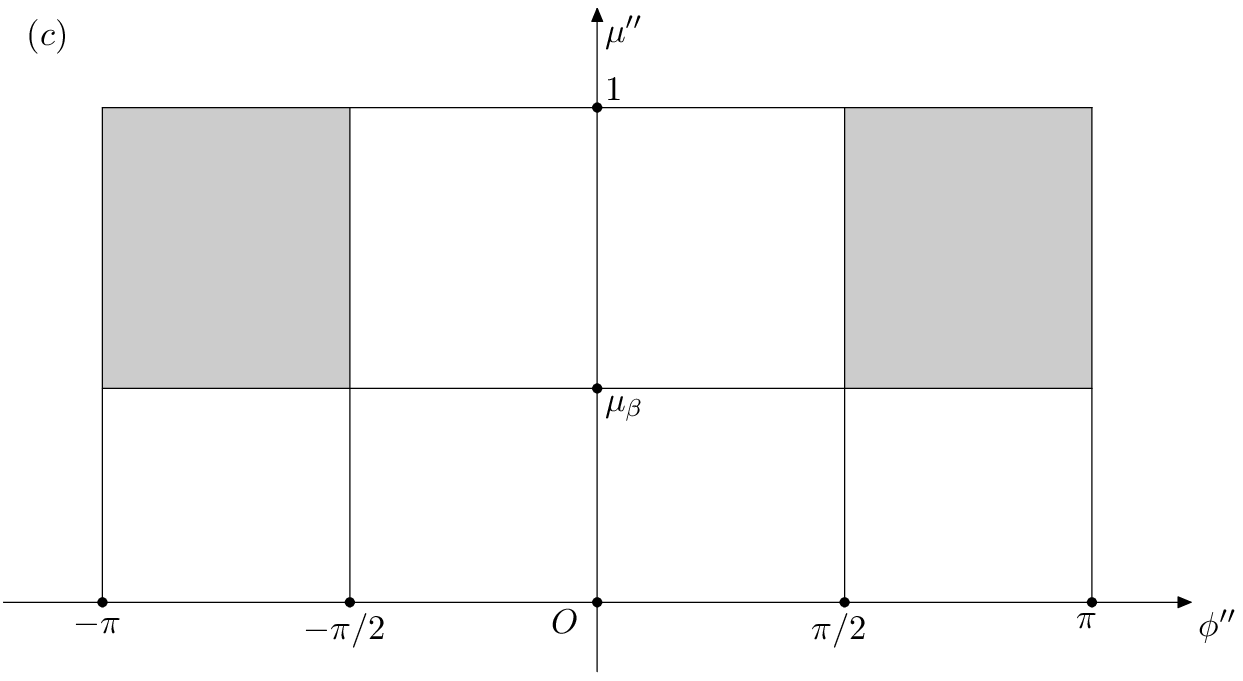} 
\caption{The same as Figure \ref{fig4}, but for the corona with a sphere geometry. The effective grey regions for panels from top to bottom correspond to the three geometrical configurations shown in Figure \ref{fig2}. The blue boundary curve in the top panel is given by Equation \eqref{eqapz2}.}
\label{fig3}
\end{figure}

The procedure for the sphere case is similar to that of the cylinder case, but the local triad at $\mathbf{r}_{\rm e}$ is constructed in a different way by (see Figure \ref{fig2})
\begin{eqnarray}\label{eqa6}
\begin{split} \left. 
\begin{array}{ll}
  \mathbf{e}'_x = - \mathbf{e}_z, \quad\mathbf{e}'_z = - \hat{\mathbf{r}}_{\rm e}, \quad 
  \mathbf{e}'_y = \mathbf{e}'_z \times \mathbf{e}'_x.
\end{array} \right.
\end{split}
\end{eqnarray}  
In addition, we should construct another triad at $\mathbf{r}_{\rm e}$ defined through
\begin{eqnarray}\label{eqa7}
\begin{split} \left\{ 
\begin{array}{ll}
  \mathbf{e}''_x &= \cos\alpha \mathbf{e}'_x + \sin\alpha \mathbf{e}'_z,\\
  \mathbf{e}''_z &= -\sin\alpha \mathbf{e}'_x + \cos\alpha \mathbf{e}'_z,\\
  \mathbf{e}''_y &=  \mathbf{e}'_y,
\end{array} \right.
\end{split}
\end{eqnarray} 
where the definition of $\alpha$ is shown in Figure \ref{fig2}. In this triad frame, all of the effective directions $(\mu'', \phi'')$ also form a grey region in the $\mu''$-$\phi''$ plane (see, Figure. \ref{fig3}). Then we can draw an effective direction $(\mu'', \phi'')$ isotropically in the rectangle region,
$[-\pi, \pi]\times[\mu_\beta, 1]$, by the following algorithm (for case (a))
\begin{eqnarray}
\begin{split}
   &{\rm do} \\
   &\quad\quad\mu'' = \mu_\beta + (1 - \mu_\beta)\xi_1, \quad
   \phi'' = -\pi + 2\pi \xi_2,\\
   &\quad\quad{\rm if( |\phi''|>\phi_1 )exit}\\ 
   &\quad\quad{\rm else\,\,\,if( \mu''>\mu''(\phi'') )exit}\\  
   &{\rm end\,\, do},
\end{split}
\end{eqnarray}
where $\mu_\beta = \cos\beta = \sqrt{1 - R^2_{\rm sp} / (R^2_e + h^2)}$, and $h = OH$ is the height of the sphere 
center of the corona above the disc, and
\begin{eqnarray}
\begin{split}
   \phi_1 = \arccos\left( \tan\alpha\cot\beta \right).
\end{split}
\end{eqnarray}
A direction $(\mu'', \phi'')$ will be accepted if it falls into 
the grey region, otherwise it will be rejected.
For case (b) 
\begin{eqnarray}
\begin{split}
   \mu'' &= \mu_\beta + (1 - \mu_\beta)\xi_1, \quad
   \phi'' = -\pi + 2\pi \xi_2.
\end{split}
\end{eqnarray}
For case (c) 
\begin{eqnarray}
\begin{split}
    &\mu'' = \mu_\beta + (1 - \mu_\beta)\xi_1, \quad
   \phi'' = \pi/2 + \pi \xi_2,\\
   &{\rm if(\phi''>\pi)\phi'' = \phi''- 2\pi}.
\end{split}
\end{eqnarray} 
With $(\mu'', \phi'')$, the initial direction of a photon can be similarly written as 
\begin{eqnarray}
\begin{split}
   \hat{\mathbf{p}}_e &= p''_x \mathbf{e}''_x + p''_y \mathbf{e}''_y + p''_z \mathbf{e}''_z 
   = p_x \mathbf{e}_x + p_y \mathbf{e}_y + p_z \mathbf{e}_z,
\end{split}\label{eqa1}
\end{eqnarray}
where 
\begin{eqnarray}\label{eqapz1}
\begin{split}\left\{ 
\begin{array}{ll}
   p''_x &= \sqrt{1 - \mu''^2}\cos\phi'', \\
   p''_y &= \sqrt{1 - \mu''^2}\sin\phi'', \\
   p''_z &= \mu''.
\end{array} \right.
\end{split}
\end{eqnarray}
From Equations \eqref{eqa6}, \eqref{eqa7}, \eqref{eqa1}, one can obtain that
\begin{equation} \label{eqapz}
\begin{split}\left\{ 
\begin{array}{ll}
   p_x &= -(\sin\alpha\cos\phi_{\rm e} p''_x - \sin\phi_{\rm e} p''_y + \cos\alpha\cos\phi_{\rm e} p''_z), \\
   p_y &= -(\sin\alpha\sin\phi_{\rm e} p''_x + \cos\phi_{\rm e} p''_y + \cos\alpha\sin\phi_{\rm e} p''_z), \\
   p_z &= -\cos\alpha p''_x + \sin\alpha p''_z.
\end{array} \right.
\end{split}
\end{equation}

In panel $(a)$ of Figure \ref{fig3}, there is a boundary curve plotted with blue color. The expression of this curve can be simply derived. From Figure \ref{fig1}, one can see that a photon reaching the corona sphere must satisfy the condition that $p_z\ge 0$ and $p_z=0$ yields the boundary curve. Using the expressions of $p_z$ given by Equation \eqref{eqapz} and $p''_x$, $p''_z$ given by Equation \eqref{eqapz1}, from $p_z=0$, we can obtain the function of the curve
\begin{eqnarray}\label{eqapz2}
\begin{split}
   \mu''(\phi'')=\frac{\cos\alpha\cos\phi''}{\sqrt{1-\cos^2\alpha\sin^2\phi''}}.
\end{split}
\end{eqnarray}

Finally, we need to update the weight $w_{\rm ini}$ by using the areas $S(R_e)$ of the grey region in Figure \ref{fig3} and the analytical expressions of $S(R_e)$ for three panels can be obtained as
\begin{eqnarray}\label{deltaSmu1}
   \begin{split} 
      S(R_e) =
        \left\{\begin{array}{lll}
           &\displaystyle 2\pi(1-\mu_\beta)-2 \arcsin(\cos\alpha\sin\phi_1)+2\mu_\beta\phi_1, 
                 \quad\quad & (a),\\ 
           &\displaystyle 2\pi(1-\mu_\beta), \quad\quad & (b),\\
           &\displaystyle \pi(1-\mu_\beta), \quad\quad &  (c).
        \end{array}\right.
   \end{split}
\end{eqnarray}

\subsection{Scattering distance sampling}

As a photon is generated, we then discuss how to determine the distance between any two scattering
points randomly. As discussed in \citep{2021ApJS..254...29X}, the scattering distance is a random variable, and its probability density distribution function (PDF) is given by
\begin{equation}\label{ps}
\begin{split}
   p(s)ds = \frac{1}{N}\exp\left(-\int_0^s \sigma_a(s') n_e(s') ds'\right)\sigma_a(s) n_e(s) ds,
\end{split}
\end{equation}
where $\sigma_a(s)=\sigma_a[h\nu, T_e(s)]$ is the averaged scattering cross section \citep{Hua1997}, $T_e(s)$ and 
$n_e(s)$ are the temperature and number density of hot electrons at $s$, respectively, $N$ is the normalization factor and
\begin{equation} 
\begin{split}
   N = 1 - \exp\left(-\int_0^{s_m} \sigma_a(s') n_e(s') ds'\right),
\end{split}
\end{equation}
where $s_m$ is the maximum distance that the path can extend in the corona region. Usually, $s_m$ can be obtained by solving an algebraic equation derived from the geometrical conditions, i.e., with the provided vectors $\hat{\mathbf{p}}_{\rm ini}, \mathbf{r}_{\rm ini}$ of the initial position and momentum direction, we have $\mathbf{r}_{\rm ini}+\hat{\mathbf{p}}_{\rm ini}s_m=\mathbf{r}_b$, where $\mathbf{r}_b$ is a position vector that falls on the boundary of the corona. Taking the square for both sides, we get a quadratic equation of $s_m$ as
\begin{equation} 
\begin{split}
   s_m^2 + 2s_m \mathbf{r}_{\rm ini}\cdot\hat{\mathbf{p}}_{\rm ini} + r^2_{\rm ini}=r^2_b.
\end{split}
\end{equation}  
In our model, we adopt the assumption that both $n_e$ and $kT_e(s)$ are constants in the corona, then the above Equation becomes 
\begin{equation} 
\begin{split}
   p(s)ds = \frac{1}{N}\exp\left(- \sigma_a n_e s\right)\sigma_a n_e ds,
\end{split}
\end{equation} 
where $N=1-\exp(-\sigma_a n_e s_m)$, which can be sampled by the inverse cumulated distribution function (CDF) method directly. That is 
\begin{equation} 
\begin{split}
   s = - \frac{1}{ n_e \sigma_a}\ln( 1 - N\xi ).
\end{split}
\end{equation}
When the scattering distance is determined, the weight $w_{\rm ini}$ should be updated as well by multiplying the normalization factor $N$, i.e., $w_{\rm ini}' = w_{\rm ini}\cdot N$.

\subsection{Scattering sampling}

In our model, we will consider inverse Comptonization of the photon in the sphere corona with 
the electrons assigned with various distribution functions.
As discussed in \cite{2021ApJS..254...29X}, Lemon can incorporate any kind of scattering readily.
Here, we mainly consider three kinds of electron distributions, i.e., the relativistic thermal,
the $\kappa$ and the power law  distributions, which are respectively given by 
\citep{2006PPCF...48..203X, 2016ApJ...822...34P}
\begin{eqnarray}\label{eqpgama} 
   \frac{dn_e}{d\gamma d\Omega}=\left\{ \begin{array}{ll}
   &\displaystyle  \frac{n_e}{4\pi} 
   \frac{ \gamma\sqrt{\gamma^2-1} }{\Theta K_2(\Theta)}\exp(-\gamma/\Theta), \\ 
   &\displaystyle  \frac{n_e}{4\pi}\frac{\gamma\sqrt{\gamma^2-1}}{N_\kappa}
   \left(1+\frac{\gamma -1 }{\kappa w}\right)^{-(\kappa+1)}, \\
   &\displaystyle  \frac{n_e}{4\pi}\frac{1}{N_p}\gamma^{-\alpha_e}, 
    \end{array}\right. 
\end{eqnarray} 
where $\gamma=1/\sqrt{1-v^2/c^2}$ is the Lorentz factor, $\gamma_1\le \gamma \le \gamma_2$, 
$d\Omega=\sin\theta d\theta d\phi$ is the solid angle,
$\Theta=kT/(m_ec^2)$ is the dimensionless temperature, $\kappa, w$ are free parameters, 
$\gamma_1,\,\, \gamma_2$ are boundary values and $N_\kappa, \,\,N_p$ are the normalization factors.
In general, we assume that $\gamma$ of these distributions is confined between $\gamma_1$ and $\gamma_2$.
$N_p$ can be written as $N_p=(\gamma_1^{1-\alpha_e}-\gamma_2^{1-\alpha_e})/(\alpha_e-1)$, while $N_\kappa$
will be evaluated numerically since its analytical expression involves special functions and 
is quite complicated \citep{2016ApJ...822...34P}. 
In \cite{2021ApJS..254...29X}, we have discussed how to sample $dn_e/(d\gamma d\Omega)$ by a method proposed by \cite{Hua1997}. Here we present a new method to deal with these two PDFs in a unified way. This method is a combination of inverse 
CDF and rejection method. For the thermal distribution, an auxiliary function is introduced as 
\begin{equation} 
\begin{split} 
   f_1(\gamma) &= \frac{1}{A}\gamma^2\exp(-\gamma/\Theta), 
\end{split}
\end{equation}
where $A$ is the normalization factor. Then the procedure of the method is given by \citep{Schnittman2013}
\begin{eqnarray}
\begin{split}
   &\textcircled{1}{\rm get\,\,a\,\,\gamma_{f_1}\,\,by\,\,sampling\,\,f_1(\gamma)},\\
   &\textcircled{2}{\rm  if \left( \gamma_{f_1}\xi \le  \sqrt{\gamma_{f_1}^2-1} \right)  accept\,\, \gamma_{f_1} }\\ 
   &{\rm \quad else \,\,\,goto\,\,\textcircled{1}}.
\end{split}
\end{eqnarray}
The algorithm of sampling $\gamma_{f_1}$ from $f_1(\gamma)$ is the inverse 
CDF method, which is equivalent to solve the following algebraic equation \citep{Schnittman2013} 
\begin{equation}\label{eqroot}
\begin{split}
    g(u)=g(u_1)-\xi[g(u_1)-g(u_2)],
\end{split}
\end{equation}
where $u = \gamma/\Theta$, $g(u) = (u^2+2u+2)\exp(-u)$, $u_1=\gamma_1/\Theta$ and $u_2=\gamma_2/\Theta$. 
It turns out that this equation can be solved by iterative method numerically. To accomplish it, we rewrite the above equation as
\begin{equation} 
\begin{split}
    u=\ln(u^2+2u+2)-\ln G_0,
\end{split}
\end{equation}
where $G_0 = g(u_1)-\xi[g(u_1)-g(u_2)]$. Then the root $u_0$ of equation. \eqref{eqroot} can be obtained by the
following algorithm:
\begin{eqnarray}
\begin{split}
   &x_1 = 0, \\
   &{\rm do }\\
   &\quad\quad  { x_2 = \ln(x_1^2+2x_1+2)-\ln G_0}\\
   &\quad\quad  {\rm if}(|x_2-x_1|<\epsilon){\rm exit}\\
   &\quad\quad  { x_1 = x_2}\\ 
   &{\rm  end\,\,do } 
\end{split}
\end{eqnarray}
and $u_0 \approx x_2$, where $\epsilon$ is the tolerance value for the accuracy of the root. 
Usually, we set it to be $\epsilon=10^{-10}$.
Once the root $u_0$ is obtained, the sample of the Lorentz factor is given by $\gamma_{f_1} = u_0\Theta$.

For the $\kappa$ distribution, the algorithm is similar and the auxiliary function is given by
\begin{equation} 
\begin{split} 
   f_\kappa(\gamma) &= \frac{1}{A} \frac{\gamma^2}{(\gamma + a)^{\kappa+1}},
\end{split}
\end{equation}
where $a=\kappa w -1$ and $A$ is the normalization factor.
Also one needs to solve the following equation to get a trial sample of $\gamma$,
\begin{equation}\label{eqroot2}
\begin{split} 
   \gamma = \left[\frac{1}{d_1G_0}(a_1\gamma^2+b_1\gamma+c_1)\right]^{1/\kappa}-a,
\end{split}
\end{equation}
where $a_1 =\kappa(\kappa-1)$, $b_1=2a\kappa$, $c_1=2a^2$, $d_1=\kappa(\kappa-1)(\kappa-2)$ and
$G_0=g(\gamma_1)-\xi[g(\gamma_1)-g(\gamma_2)]$, the function $g(\gamma)$ is given by
\begin{eqnarray} 
\begin{split}
   g(\gamma) = \frac{a_1\gamma^2+b_1\gamma+c_1 }{d_1(\gamma+a)^\kappa}.
\end{split}
\end{eqnarray}
Equation \eqref{eqroot2} can be numerically solved by iterative method as well and the algorithm is 
similar 
\begin{eqnarray}
\begin{split}
   &x_1 = 1, \\
   &{\rm do }\\
   &\quad\quad  { x_2 = \left[\frac{1}{d_1G_0}(a_1x_1^2+b_1x_1+c_1)\right]^{1/\kappa}-a}\\
   &\quad\quad  {\rm if}(|x_2-x_1|<\epsilon){\rm exit}\\
   &\quad\quad  { x_1 = x_2}\\ 
   &{\rm end\,\,do } 
\end{split}
\end{eqnarray}
and $\gamma_{f_\kappa} \approx x_2$. For the power law distribution, the samples of $\gamma$ can be 
obtained directly by inverse CDF method as
\begin{eqnarray} 
\begin{split}
    \gamma = \left[\gamma_1^{1-\alpha_e}(1-\xi)+\gamma_2^{1-\alpha_e})\right]^{\frac{1}{1-\alpha_e}}.
\end{split}
\end{eqnarray}

The momentum direction (MD) $(\theta, \phi)$ of the electron can be drawn isotropically as follows
\begin{eqnarray} 
\begin{split}
    \mu = -1 + 2\xi_1, \quad \phi = 2\pi \xi_2,
\end{split}
\end{eqnarray}
where $\mu=\cos\theta$. Then we calculate the ratio of the total cross section of Klein-Nishina and Thomson by \citep{Hua1997}
\begin{equation} 
\begin{split}
  \frac{\sigma_{\rm KN}}{\sigma_T} = \frac{3}{4\epsilon}\left[ \left( 1 - \frac{4}{\epsilon} - 
  \frac{8}{\epsilon^2} \right) \ln( 1 + \epsilon )
      + \frac{1}{2} + \frac{8}{\epsilon}  - \frac{1}{2(1+\epsilon)^2}\right] ,
\end{split}
\end{equation}
where 
\begin{equation} 
\begin{split}
    \epsilon = \frac{ 2 h\nu }{m_ec^2} \gamma ( 1 - v \mu ),
\end{split}
\end{equation}
and $h\nu$ is the energy of the incident photon, $v = \sqrt{1-1/\gamma^2}$ is speed of the electron.
A random number $\xi$ is generated to determine the acceptence of $\gamma, \mu, \phi$ as the Lorentz factor 
and velocity direction of the scattering electron. If $\xi \le \sigma_{\rm KN}/\sigma_T$ 
they are accepted, otherwise rejected.

\subsection{Lorentz Transformation of Stokes parameters}
The polarization states are described by the Stokes parameters (SPs): $\mathbf{S}=(I, Q, U, V)^T$ and the polarization 
vector (PV) $\mathbf{f}$.
The Compton scattering will inevitably change the polarization states of the photons. In this subsection we will discuss how to describe and trace these changes in Lemon in a detailed way. These discussions however can be found in, e.g., \cite{2012ApJ...744...30K}. For the purpose of completeness, we shall give these descriptions in a more consistent way as follows.

1. As the MD $\mathbf{k}$ of the incident photon is given, we first construct the photon triad as
$\mathbf{e}_z(p)=\mathbf{k},\,\mathbf{e}_x(p)=\mathbf{f},\,\mathbf{e}_y(p)=\mathbf{e}_z(p)\times\mathbf{e}_x(p)$. For unpolarized radiations, the basis vector $\mathbf{e}_x(p)$ is set as $\mathbf{e}_x(p)=\mathbf{k}\times\mathbf{z}/|\mathbf{k}\times\mathbf{z}|$,  where $\mathbf{z}=(0, 0, 1)$.

2. Then in the photon triad $\mathbf{e}_i(p)$ we obtain the MD $\mathbf{p}_e$ and Lorentz factor $\gamma_e$ of the scattering electron by sampling the distribution function given by Equation \eqref{eqpgama} as discussed in the above subsection. Equivalently, $\mathbf{p}_e$ can be expressed by $(\mu_e, \phi_e)$, i.e.,
$\mathbf{p}_e = \sqrt{1-\mu_e^2}\cos\phi_e\mathbf{e}_x(p) + \sqrt{1-\mu_e^2}\sin\phi_e\mathbf{e}_y(p) +  \mu_e \mathbf{e}_z(p)$.
With $\mathbf{p}_e$ we can construct the static electron triad with respect to the photon triad $\mathbf{e}_i(p)$ as $\mathbf{e}_z(e)=\mathbf{p}_e,\,\mathbf{e}_y(e)=\mathbf{k}\times\mathbf{p}_e/|\mathbf{k}\times\mathbf{p}_e|,\,\mathbf{e}_x(e)=\mathbf{e}_y(e)\times\mathbf{e}_z(e)$. We denote the MD of the incident photon as $(\mu_e, \phi_e)$ in this triad.

3. To complete the Compton scattering, we need to transform the SPs into the rest frame of the electron. To do so, we should first carry out a rotation and get the SPs defined with respect to the electron triad $\mathbf{e}_i(e)$, that is $\mathbf{S}_e=\mathbf{M}(\phi_e)\mathbf{S}$, where
\begin{equation} 
\begin{split}
    \mathbf{M}(\phi_e)=\left(
       \begin{array}{cccc}
          1 & 0 & 0 & 0 \\
          0 &  \cos2\phi_e & \sin2\phi_e & 0 \\
          0 & -\sin2\phi_e & \cos2\phi_e & 0 \\
          0 & 0 & 0 & 1
       \end{array}
    \right)
\end{split}
\end{equation}
is the rotation matrix (see \cite{1960ratr.bookC}). At the same time the PV $\mathbf{f}$ has been rotated into the plane of $\mathbf{p}_e$ and $\mathbf{k}$ as well.
As demonstrated by \cite{2012ApJ...744...30K}, $\mathbf{S}_e$ will keep invariant under the Lorentz transformation, which means that we can get the SPs in the rest frame of the electron directly as: $\widetilde{\mathbf{S}}_e=\mathbf{S}_e$ (from now on, the quantities in the rest frame of the electron are denoted with a tilde $\tilde{}$). The MD $(\mu_e, \phi_e)$ and frequency $\nu$ of the incident photon will be transformed as:
\begin{equation}\label{lorentz_1}
\begin{split}
   \widetilde{\mu}_e &= \frac{\mu_e-v_e}{1-\mu_ev_e},\,\,\widetilde{\phi}_e=\phi_e,\\
   \widetilde{\nu} &= \nu\gamma(1-\mu_ev_e),
\end{split}
\end{equation}
where $v_e$ is the velocity of the electron.

4. With $(\widetilde{\mu}_e, \widetilde{\phi}_e)$ we could reconstruct the MD $\widetilde{\mathbf{k}}$ of the photon in the rest frame of the electron. Using $\widetilde{\mathbf{k}}$, we can construct another photon triad with respect to the electron rest frame as: $\widetilde{\mathbf{e}}_z(p)=\widetilde{\mathbf{k}},\,\widetilde{\mathbf{e}}_y(p)=\widetilde{\mathbf{k}}\times\widetilde{\mathbf{z}}/|\widetilde{\mathbf{k}}\times\widetilde{\mathbf{z}}|,\,\widetilde{\mathbf{e}}_x(p)=\widetilde{\mathbf{e}}_y(p)\times\widetilde{\mathbf{e}}_z(p)$, where $\widetilde{\mathbf{z}}=(0,0,1)$ is the $z$-axis of the electron frame. In frame $\widetilde{\mathbf{e}}_i(p)$, we redenote the MD of the electron as
$\widetilde{\mathbf{p}}_e$.

5. Simulating the Compton scattering in the rest frame of the electron, we sample the Klein-Nishina differential cross section $d\sigma_{\rm KN}/(d\widetilde{\mu}'d\widetilde{\phi}')$ to get the scattered frequency $\widetilde{\nu}'$ and MD $(\widetilde{\mu}'_e, \widetilde{\phi}'_e)$ (see, e.g., \cite{1983ASPRv...2..189P} and \cite{Hua1997}), which is defined with respect to the triad $\widetilde{\mathbf{e}}_z(p)$ and
\begin{equation} 
\begin{split}
   \widetilde{\nu}'=\frac{\widetilde{\nu}}{\displaystyle 1+ h\widetilde{\nu}/(m_ec^2)(1-\widetilde{\mu}'_e)}.
\end{split}
\end{equation} 
From $(\widetilde{\mu}'_e, \widetilde{\phi}'_e)$ we can get the scattered MD vector $\widetilde{\mathbf{k}}'$ and use it to construct the scattered photon triad as $\widetilde{\mathbf{e}}'_z(p)=\widetilde{\mathbf{k}}',\,\widetilde{\mathbf{e}}'_y(p)=\widetilde{\mathbf{k}}'\times\widetilde{\mathbf{k}}/|\widetilde{\mathbf{k}}'\times\widetilde{\mathbf{k}}|, \,\widetilde{\mathbf{e}}'_x(p)=\widetilde{\mathbf{e}}'_y(p)\times\widetilde{\mathbf{e}}'_z(p)$.

Then we carry out a rotation given by $\widetilde{\mathbf{S}}_{p}=\mathbf{M}(\widetilde{\phi}'_e)\widetilde{\mathbf{S}}_e$ to get the SPs defined with respect to the scattering plane determined by $\widetilde{\mathbf{k}}'$ and $\widetilde{\mathbf{k}}$. From $\widetilde{\mathbf{S}}_{p}$, the scattered SPs $\widetilde{\mathbf{S}}'_{p}$ can be obtained by the Fano's Matrix as \citep{Fano1949, Fano1957, McMa1961}:
\begin{equation} 
\begin{split}
   \widetilde{\mathbf{S}}'_{p}=\mathbf{F}(\widetilde{\nu},\widetilde{\nu}',\widetilde{\theta}'_e)\widetilde{\mathbf{S}}_{p},
\end{split}
\end{equation}
where
\begin{equation} 
\begin{split}
    \mathbf{F}(\widetilde{\nu},\widetilde{\nu}',\widetilde{\theta}'_e) =\left(\frac{\widetilde{\nu}'}{\widetilde{\nu}}\right)^2\left(
       \begin{array}{cccc}
          F_{0} & F_3 & 0 & 0 \\
          F_3 &  F_{33} & 0 & 0 \\
          0 & 0 & F_{11} & 0 \\
          0 & 0 & 0 & F_{22}
       \end{array}
    \right)
\end{split}
\end{equation}
and 
\begin{equation} 
\begin{split}
F_0 &= \displaystyle\frac{\widetilde{\nu}'}{\widetilde{\nu}}+ \frac{\widetilde{\nu}}{\widetilde{\nu}'}-
\sin^2\widetilde{\theta}'_e,\quad F_3 = \sin^2\widetilde{\theta}'_e,\quad F_{11}=2\cos\widetilde{\theta}'_e,\\
F_{22} &= \left(\frac{\widetilde{\nu}'}{\widetilde{\nu}}+ \frac{\widetilde{\nu}}{\widetilde{\nu}'}\right)
\cos\widetilde{\theta}'_e,\quad F_{33} = 1+\cos^2\widetilde{\theta}'_e.
\end{split}
\end{equation}

6. Now we construct a triad by using $\widetilde{\mathbf{k}}'$ and $\widetilde{\mathbf{p}}_e$, i.e., 
$\widetilde{\mathbf{e}}''_z(p)=\widetilde{\mathbf{k}}',\,\widetilde{\mathbf{e}}''_y(p)=\widetilde{\mathbf{k}}'\times\widetilde{\mathbf{p}}_e/
|\widetilde{\mathbf{k}}'\times\widetilde{\mathbf{p}}_e|, \,\widetilde{\mathbf{e}}''_x(p)=\widetilde{\mathbf{e}}''_y(p)\times\widetilde{\mathbf{e}}''_z(p)$. We then can obtain the angle between the plane $\widetilde{\mathbf{k}}'$-$\widetilde{\mathbf{p}}_e$ and plane $\widetilde{\mathbf{k}}'$-$\widetilde{\mathbf{k}}$ (or equivalently the angle between basis vector $\widetilde{\mathbf{e}}'_x(p)$ and $\widetilde{\mathbf{e}}''_x(p)$) as (refer to Equation. (8) of \cite{2012ApJ...744...30K}):
\begin{equation}\label{phi0def}
\begin{split}
   \phi_0=-{\rm sign}[\widetilde{\mathbf{e}}'_x(p)\cdot\widetilde{\mathbf{e}}''_y(p)]
   \arccos[\widetilde{\mathbf{e}}'_y(p)\cdot\widetilde{\mathbf{e}}''_y(p)].
\end{split}
\end{equation}
With $\phi_0$ we can do a rotation to get the SPs defined with respect to the triad $\widetilde{\mathbf{e}}''_i(p)$ as
$\widetilde{\mathbf{S}}''_{p}=\mathbf{M}(\phi_0)\widetilde{\mathbf{S}}'_{p}$, which could be transformed back into the static frame directly.

7. At this stage, the Compton scattering has been completed and we implement another Lorentz transformation to bring all the quantities back into the static reference. For the scattered MD $\widetilde{\mathbf{k}}'$ we first transform it from the $\widetilde{\mathbf{e}}_y(p)$ frame to the electron rest frame through
\begin{equation} 
\begin{split}
   \widetilde{k}'^m(e) = \widetilde{k}'^i(p)\widetilde{e}_{(i)}^m(p),
\end{split}
\end{equation}
where $\widetilde{e}_{(i)}^m(p)$ are the components of the $i$-th basis vector $\widetilde{\mathbf{e}}_i(p)$ with respect to the electron rest frame and $\widetilde{k}'^m(p), \widetilde{k}'^i(e)$ are the components of $\widetilde{\mathbf{k}}'$ with respect to the photon and electron triad respectively. Then the Lorentz transformation of $\widetilde{\nu}'$ and $\widetilde{k}'^m(e)$ can be 
written as
\begin{equation} 
\begin{split}
   &\nu'  = \widetilde{\nu}'\gamma[1+\widetilde{k}'^z(e)v_e],\,\,
   k'^z(e) = \frac{\widetilde{k}'^z(e) + v_e}{1+\widetilde{k}'^z(e)v_e},\\
   &k'^x(e) = \frac{\widetilde{k}'^x(e)}{\gamma[1+\widetilde{k}'^z(e)v_e]},\,\,
   k'^y(e) = \frac{\widetilde{k}'^z(e)}{\gamma[1+\widetilde{k}'^z(e)v_e]}.
\end{split}
\end{equation}
The scattered direction in the static frame can be obtained from
\begin{equation} 
\begin{split}
   k'^m = k'^i(e)e_{(i)}^j(e)e_{(j)}^m(p),
\end{split}
\end{equation}
and the scattered vector $\mathbf{k}'=k'^m\mathbf{e}_m$.
The scattered SPs in the static frame are given by $\widetilde{\mathbf{S}}''_{p}$ and the scattered polarization vector $\mathbf{f}'$ is obviously in the $\mathbf{p}_e$-$\mathbf{k}'$ plane and can be expressed as 
$\mathbf{f}'=[\mathbf{p}_e-(\mathbf{p}_e\cdot\mathbf{k}')\mathbf{k}']/N$, where $N$ is the normalization factor. As $\mathbf{k}'$, $\widetilde{\mathbf{S}}''_{p}$ and $\mathbf{f}'$ are obtained, one can continue to trace the next transfer and scattering of the photon. Simultaneously, one can record the contributions of the photon made to the observed quantities at the scattering point, which will be discussed in the next section.

\subsection{Spectrum and polarization estimation}
Lemon used a scheme that can improve the efficiency and accuracy of spectrum and polarization evaluations, since the information at 
any scattering point can perform contributions to the observed spectrum and polarization. Under the Neumann expansion
solution of differential-integral equation, the scheme is equal to the introduction of a 
$\delta$-function and recording function (for more detailed discussions, refer to \cite{2021ApJS..254...29X}). 
This scheme has actually been applied widely and implemented in many codes dealing with Ly$\alpha$ radiative transfer
(e.g., see \cite{2022ApJS..259....3S}, where the scheme is named as “peeling-off technique”, also
known as “next event estimation” or “shadow rays” 
\citep{1984ApJ...278..186Y, 2007ApJ...657L..69L, 2012MNRAS.424..884Y}, one may also refer to
\cite{whit2011} and \cite{noeb2019} for reviews). We will 
use this scheme to reduce the noise generated by the Monte Carlo method and obtain the results with 
high signal-to-noise ratio.

Now we discuss the specific procedures of the estimation scheme using the conventions and triads established in the last subsection. The observer is assumed to be located at a direction $\mathbf{n}_{\rm obs}(\mu_{\rm obs}, \phi_{\rm obs})$. Due to the axial symmetry of the system, we can choose $\phi_{\rm obs}$ randomly, i.e., $\phi_{\rm obs}=2\pi\xi$. Then using the coefficients of the photon and electron triad, we can transform $\mathbf{n}_{\rm obs}$ into the two triads directly by
\begin{equation} 
\begin{split}
   n_{\rm obs}^{(m)}(p) &= n_{\rm obs}^i e_{i}^{(m)}(p),\\
   n_{\rm obs}^{(m)}(e) &= n_{\rm obs}^i e_{i}^{(j)}(p)e_{j}^{(m)}(e),
\end{split}
\end{equation}
where $e_{i}^{(m)}$ is the inverse of matrix $e_{(m)}^{i}$. Since all of the triads are orthonormal, we have $e_{i}^{(m)}e_{(m)}^{j}=\delta_i^j$, $e_{i}^{(m)}e_{(n)}^{i}=\delta_n^m$ and $e_{i}^{(m)}={e^T}_{(m)}^{i}$, where the superscript $T$ represents the matrix transpose. Through a Lorentz transformation, we get the observer direction in the rest frame of the electron as
\begin{equation} 
\begin{split} 
   \widetilde{n}_{\rm obs}^{(x)}(e) &=\frac{n_{\rm obs}^{(z)}(e)}{\gamma D},\,\,
   \widetilde{n}_{\rm obs}^{(y)}(e) =\frac{n_{\rm obs}^{(z)}(e)}{\gamma D},\\
   \widetilde{n}_{\rm obs}^{(z)}(e) &=\frac{n_{\rm obs}^{(z)}(e)-v_e}{D}.
\end{split}
\end{equation}
where $D=1-n_{\rm obs}^{(z)}(e)v_e$.
Using the triad matrix $\widetilde{e}^{i}_{(m)}(p)$ of $\widetilde{\mathbf{e}}_i(p)$, we can further transform the observer direction into the photon frame as
\begin{equation} 
\begin{split} 
   \widetilde{n}_{\rm obs}^{(m)}(p) = \widetilde{n}_{\rm obs}^{(n)}(e)\widetilde{e}_{n}^{(m)}(p),
\end{split}
\end{equation}
or in the vector form $\widetilde{\mathbf{n}}_{\rm obs}(p)=\widetilde{n}_{\rm obs}^{(m)}(p)\widetilde{\mathbf{e}}_m(p)$.
Then the cosine of the scattering angle in the rest frame of the electron is given by $\widetilde{\mu}'_e=\cos\widetilde{\theta}'_e=\widetilde{n}_{\rm obs}^{(z)}(p)$, from which we can get the scattered frequency in the electron rest and static frame respectively as
\begin{equation} 
\begin{split}  
   \widetilde{\nu}'_{\rm obs}&=\frac{\widetilde{\nu}}{\displaystyle 1+ h\widetilde{\nu}/(m_ec^2)(1-\widetilde{\mu}'_e)},\\
   \nu'_{\rm obs} &= \widetilde{\nu}'_{\rm obs}\gamma[1+\widetilde{k}'^z(e)v_e],
\end{split}
\end{equation}
where $\widetilde{\nu}$ is the frequency of the incident photon given by Equation \eqref{lorentz_1}, $m_ec^2$ is the rest energy of the electron. To get SPs for a given observer direction, we should rotate the SPs $\widetilde{\mathbf{S}}_e$ into scattering plane by $\widetilde{\mathbf{S}}_{p}^{\rm obs}=\mathbf{M}(\widetilde{\phi}'_e)\widetilde{\mathbf{S}}_e$, and obviously
\begin{equation} 
\begin{split} 
   \cos\widetilde{\phi}'_e=\frac{\widetilde{n}_{\rm obs}^{(x)}(p)}{\sin\widetilde{\theta}'_e},\quad
   \sin\widetilde{\phi}'_e=\frac{\widetilde{n}_{\rm obs}^{(y)}(p)}{\sin\widetilde{\theta}'_e}.
\end{split}
\end{equation}
Then the scattered SPs can be obtained by the Fano's Matrix as
\begin{equation} 
\begin{split}
   \widetilde{\mathbf{S}}^{\prime\rm obs}_{p}=\mathbf{F}(\widetilde{\nu},\widetilde{\nu}'_{\rm obs},\widetilde{\theta}'_e)\widetilde{\mathbf{S}}_{p}^{\rm obs}.
\end{split}
\end{equation}
With $\widetilde{\mathbf{n}}_{\rm obs}(p)$ we can construct two triads as before:
\begin{equation} 
\begin{split}
   \widetilde{\mathbf{e}}'_z(p)&=\widetilde{\mathbf{n}}_{\rm obs}(p),\,\widetilde{\mathbf{e}}'_y(p)=
\widetilde{\mathbf{n}}_{\rm obs}(p)\times\widetilde{\mathbf{k}}
/|\widetilde{\mathbf{n}}_{\rm obs}(p)\times\widetilde{\mathbf{k}}|, \\
\widetilde{\mathbf{e}}'_x(p)&=\widetilde{\mathbf{e}}'_y(p)\times\widetilde{\mathbf{e}}'_z(p),
\end{split}
\end{equation} 
and
\begin{equation} 
\begin{split}
   \widetilde{\mathbf{e}}''_z(p)&=\widetilde{\mathbf{n}}_{\rm obs}(p),\,\widetilde{\mathbf{e}}''_y(p)=
\widetilde{\mathbf{n}}_{\rm obs}(p)\times\widetilde{\mathbf{p}}_e
/|\widetilde{\mathbf{n}}_{\rm obs}(p)\times\widetilde{\mathbf{p}}_e|, \\
\widetilde{\mathbf{e}}''_x(p)&=\widetilde{\mathbf{e}}''_y(p)\times\widetilde{\mathbf{e}}''_z(p),
\end{split}
\end{equation} 
from which we can get rotation angle $\phi_0$ given by Eq. \eqref{phi0def} and then the scattered SPs defined with respect to the 
plane of $\widetilde{\mathbf{n}}_{\rm obs}(p)$ and $\widetilde{\mathbf{p}}_e$ through $\widetilde{\mathbf{S}}^{\prime\prime\rm obs}_{p}=\mathbf{M}(\phi_0)\widetilde{\mathbf{S}}'_{p}$. Finally, the polarization vector is given by $\mathbf{f}'_{\rm obs}=[\mathbf{p}_e-(\mathbf{p}_e\cdot\mathbf{n}_{\rm obs})\mathbf{n}_{\rm obs}]/N$.

As $\nu'_{\rm obs}$, $\widetilde{\mathbf{S}}^{\prime\prime\rm obs}_{p}$ and $\mathbf{f}'_{\rm obs}$ are obtained, we can eventually record the contribution made by this scattering site to the observed spectrum and polarization.. Before that, we should do a final rotation to get the SPs defined with respect to the static triad of the observer, which is constructed by
\begin{equation} 
\begin{split}
    \mathbf{e}_z(\rm obs)&= \mathbf{n}_{\rm obs},\,\mathbf{e}_x({\rm obs}) = (- n_y^{\rm obs}, n_x^{\rm obs}, 0), 
 \mathbf{e}_y({\rm obs})=\mathbf{e}_z({\rm obs})\times\mathbf{e}_x({\rm obs}).
\end{split}
\end{equation}
By the definition of $\mathbf{e}_x({\rm obs})$, we have assumed that the polarization angle is measured from the a direction parallel to the disc in the sky plane. A polarization vector along the north-south direction corresponds to a polarization angle of 90 degrees. This definition is different from the convension adopted by \cite{2022MNRAS.510.3674U} with a 90$^\circ$ rotation. The triad associated with $\mathbf{f}'_{\rm obs}$ can be obtained by
\begin{equation} 
\begin{split}
    \mathbf{e}_z(p)&= \mathbf{n}_{\rm obs},\,\mathbf{e}_x(p) = \mathbf{f}'_{\rm obs}, 
 \mathbf{e}_y(p)=\mathbf{e}_z(p)\times\mathbf{e}_x(p).
\end{split}
\end{equation}
Then the rotation angle can be calculated by
\begin{equation} 
\begin{split}
   \phi_0=-{\rm sign}[\mathbf{e}_x(p)\cdot \mathbf{e}_y({\rm obs})]
   \arccos[\mathbf{e}_y(p)\cdot\mathbf{e}_y({\rm obs})].
\end{split}
\end{equation}
With $\phi_0$, we get the SPs defined with respect to the observer triad: $\mathbf{S}^{\prime\prime\rm obs}_{p}=\mathbf{M}(\phi_0) \widetilde{\mathbf{S}}^{\prime\prime\rm obs}_{p}$. Then one can synthesize the spectrum $\nu L_\nu$, polarization degree (PD) $\delta$ and polarization angle (PA) by
\begin{equation} 
\left\{\begin{array}{rl}
    \nu L_\nu &= I_{\rm obs} = \displaystyle\sum \frac{I^{\prime\prime}_{\rm obs}w\sigma_{\rm KN}\exp(-\tau_{\rm obs})h\nu}{\Delta \ln\nu}, \\
    Q_{\rm obs} &=\displaystyle \sum \frac{Q^{\prime\prime}_{\rm obs}w\sigma_{\rm KN}\exp(-\tau_{\rm obs})h\nu}{\Delta \ln\nu}, \\
    U_{\rm obs} &=\displaystyle \sum \frac{U^{\prime\prime}_{\rm obs}w\sigma_{\rm KN}\exp(-\tau_{\rm obs})h\nu}{\Delta \ln\nu}, \\
    \delta &= \displaystyle\frac{Q_{\rm obs}}{I_{\rm obs}},\\
    PA & = \displaystyle\frac{1}{2}\arctan\left(\frac{U_{\rm obs}}{Q_{\rm obs}}\right),
\end{array}\right.
\end{equation} 
where $(I^{\prime\prime}_{\rm obs}, Q^{\prime\prime}_{\rm obs}, U^{\prime\prime}_{\rm obs})^T=\mathbf{S}^{\prime\prime\rm obs}_{p}$, $w$ is the weight, $\sigma_{\rm KN}$ is the Klein-Nishina differential cross section and $\tau_{\rm obs}$ is the optical depth from the scattering site to the boundary (e.g. \cite{2021ApJS..254...29X}).

\section{Results}
\label{Sec: results}

In this paper we mainly investigate the effects of geometries on the observed spectrum and polarization of the disc-corona system. 
The corona is assumed to be assigned with a slab, a sphere and a cylinder geometry composed of hot electron gas. 
In the following, we will present the primary results of our models.

\subsection{Example demonstrations}

\begin{figure}[t]
\center  
\includegraphics[scale = 0.5]{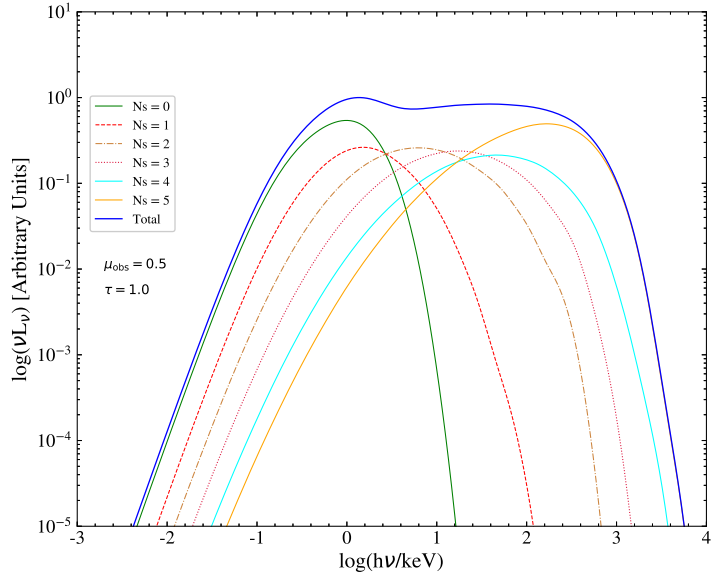} 
\caption{The spectra of a corona with spherical configuration for various scattering number. The parameters are: $R_{\rm in}$ = 6.0 $r_g$, $R_{\rm out}$ = 100.0 $r_g$, $H=0.0$ $r_g$, $\tau=1.0$, $T_e=100$ keV.}
\label{Fig:showcase1}
\end{figure}

\begin{figure}[t]
\center  
\includegraphics[scale = 0.53]{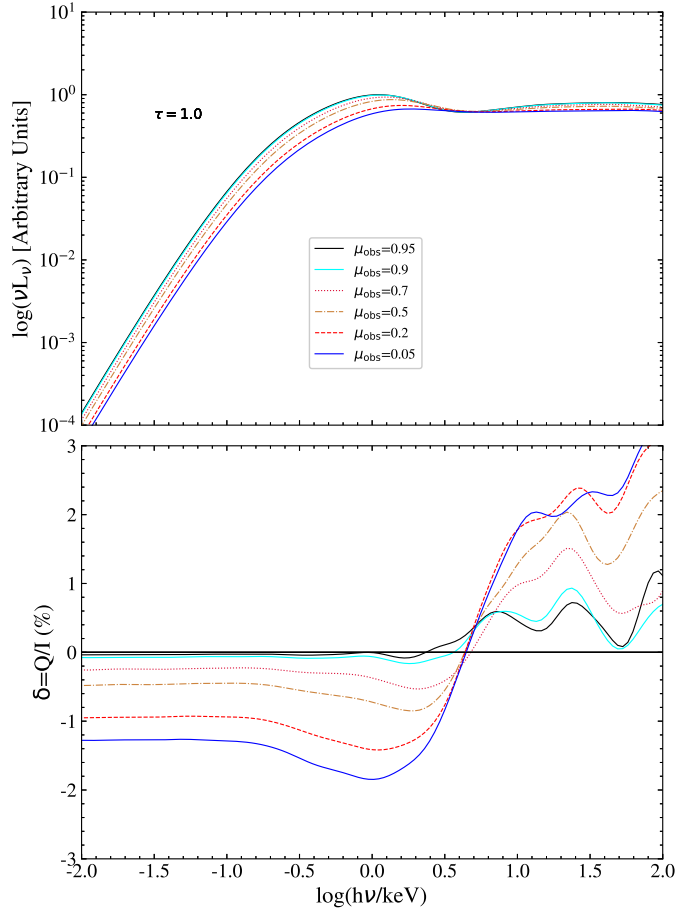} 
\caption{The same as Figure \ref{Fig:showcase1}, but for different viewing inclination angles. The top and bottom panels are the spectra and PDs, respectively.}
\label{Fig:showcase2}
\end{figure}

\begin{table}[t!]
\centering
\caption{Parameters values for all simulations.}
\begin{tabular}{ p{2cm} p{2cm} }
 \hline
 \hline
 %\multicolumn{2}{c}{ parameters values } \\
 %\hline
   parameters &  values  \\
 \hline
   $R_{\rm out}$ &  $200 $~$r_g$  \\
   $R_{\rm in}$  &  $6 \,\,r_g$   \\ 
   $\eta$        &  0.1       \\
   $\dot{m}$     &  $1.40 \times 10^{18}\rm g/s$ \\
   $M_{\rm BH}$  &  $10\,\,M_{\sun}$ \\ 
 \hline 
 \hline
\end{tabular} 
\label{table1}
\end{table}  
 
We first demonstrate the global pictures of the spectra for various viewing angles and scattering numbers. To accomplish this, the values of all other parameters must be fixed. The results are shown in Figure \ref{Fig:showcase1} and \ref{Fig:showcase2}. In Figure \ref{Fig:showcase1} we show the spectra with various scattering numbers (top panel)  viewed at  $\mu_{\rm obs}=0.5$. From the figures, one can see the typical characteristics for a Comptonized spectrum, i.e., at high energy bands, the spectrum is composed by a power law followed with a steep exponential high-energy cut-off due to the KN effect (e.g. \cite{2015MNRAS.451.4375F, 2021MNRAS.507...55M}). Also as the higher the energy is, the louder the noise is. The green line represents the spectrum formed by the multi-temperature black body radiations escaping from the disc directly. As the  scattering number increases, the corresponding spectrum becomes harder. In Figure \ref{Fig:showcase2}, we show the spectra and polarizations with respect to the inclination angles, where the curves with different colors represent different cosine of the viewing angles. From the figure one can see that as the inclination angle increases the PD $\delta$ will increase, but the intensity $I$ will decrease. This is because that photons will go through a larger optical depth and thus suffer from more scatterings at higher inclination.

\subsection{Parameter Settings} 

The parameters of the disc are set to be the same for the three kinds of coronas (see Table \ref{table1}). The inner and outer radii of the disc are set to be $R_{\rm in} = 6$ $r_g$ and $R_{\rm out} = 200$ $r_g$, respectively. The mass of the central black hole and the mass accretion rate of the disc are 10 $M_{\sun}$ and $1.4\times 10^{18}$ g/s.

The slab-like corona is composed of parallel planes sandwiching and covering the disc completely \citep{1993A...275..337P, 2010ApJ...712..908S}.  Then, the inner and outer radii of the slab are set to be the same as the disc. The height and the temperature of the slab are given by $h$ and $kT_e$. Former researches show that the dependence of results on the height $h$ is not significant (e.g. \citet{2022MNRAS.510.3674U}). Hence, in all the calculations, we will set $h$ to be $1\,\,r_g$. The Thomson optical depth $\tau$ of the slab is given as $n_e\sigma_T h$. We will simulate the cases with $\tau=0.5,\,\,1.0$ and $2.0$, respectively.

The geometry of corona with a spherical configuration is determined by its radius $R_{\rm sp}$ and height $H$. For appropriately setting values of these two parameters, the configuration of the corona can either be extending that can fully cover the disk, or be compact and located above the disc. Many evidences have shown that the size of the hot corona is most likely very small and close to the central compact \citep{2015MNRAS.451.4375F, 2022MNRAS.510.3674U} \citep{2015MNRAS.451.4375F, 2020A&A...644A.132U}, which is the well known lamppost model \citep{1996MNRAS.283..193Z, 1999MNRAS.309..561Z}. In other models, the size of the sphere can be as large as fully covering the whole disc (e.g. \cite{2018A...619A.105T}). Thus in our simulations, we will set $R_{\rm sp}$ = 10 $r_g$ for a compact configuration and $R_{\rm sp}$ = 200.0 $r_g$ for a extending configuration. The optical depth $\tau=n_e\sigma_T R_{\rm sp}$ of the sphere is measured from the center to the boundary and set to be $\tau=0.5,\,\,1.0$ and $2.0$ as well.

The corona with a cylindrical geometry is usually used to describe the outflowing materials, or a jet \citep{2004A...413..535G}. Then the corona will be assigned with a bulk velocity $\beta=v/c$. As long as the motion is not extremely relativistic, its influence on the results will not be significant \citep{2022MNRAS.510.3674U}. Thus in our simulations we can set $\beta=0$ and focus on the impact of other relevant geometrical parameters on the results. The Thomson optical depth $\tau=n_e\sigma_TR_c$ is measured along the horizontal direction and is also set to be $\tau=0.5,\,\,1.0,\,\,2.0$. The electron temperature is assigned with the value of $kT_e= 100.0$ keV for all the situations.

\subsection{Spectra and Polarizations}

With the parameter settings and assumptions given above, we carry out the simulations to study the dependencies of observed spectra and polarizations on the geometries of the corona. The results will be presented as follows.
 
%~~~~~~~~~~~~~~~~~~~~~~~~~~~~~~~~~~~~~~~~~~~~~~~~~~~~~~~~~~~~~~~~~~~~~~~~~~~~~~~~~~~~~~~~~~~
%~~~~~~~~~~~~~~~~~~~~~~~~~~~~~~~~~~~~~~~~~~~~~~~~~~~~~~~~~~~~~~~~~~~~~~~~~~~~~~~~~~~~~~~~~~~

\begin{figure*}[t]
\center 
\includegraphics[scale = 0.5]{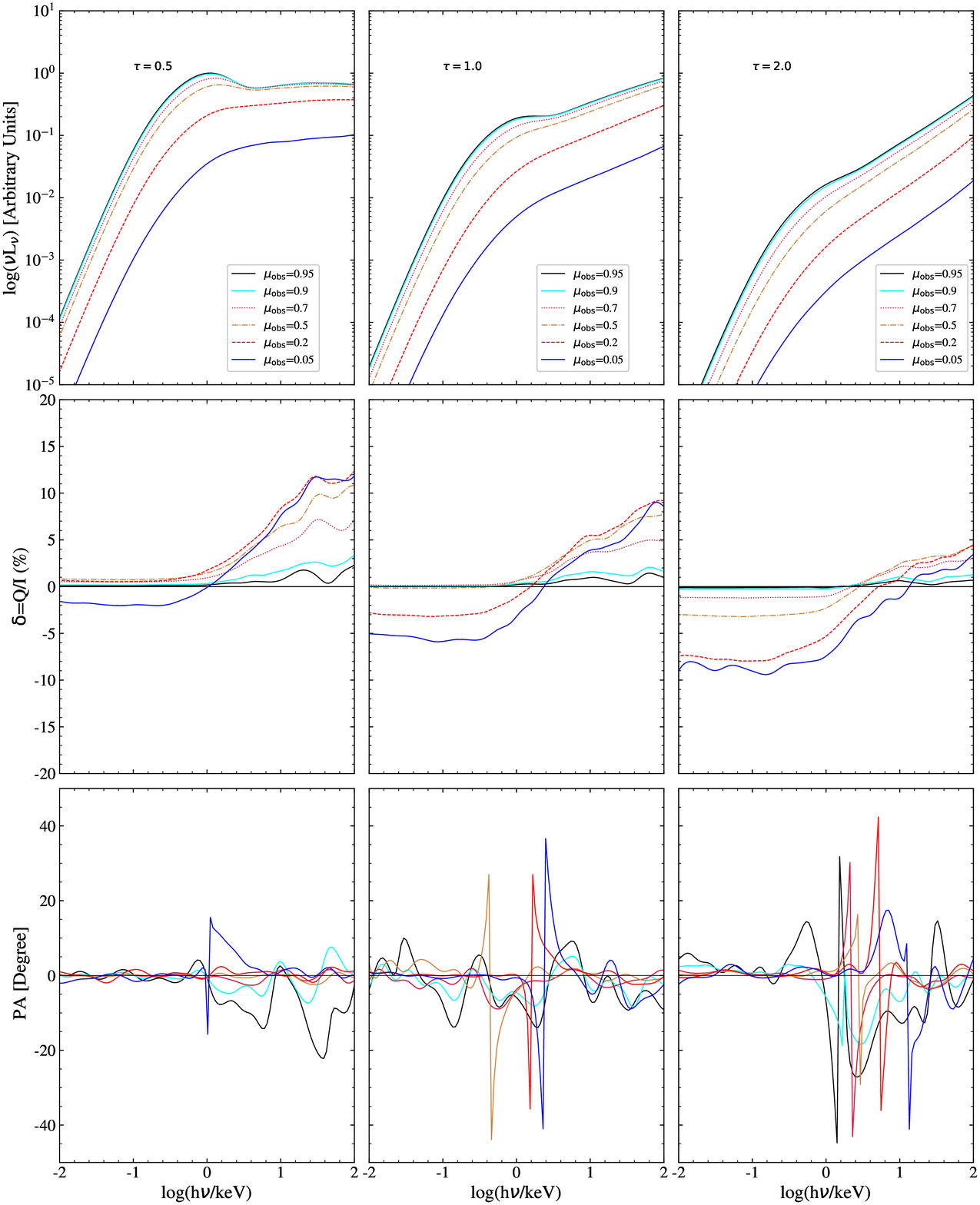} 
\caption{The synthesized spectra (top panel), PDs (middle panel) and PAs (bottom panel) of a corona with slab geometry viewing from various inclination angles. The parameters are: $T_e=100$ keV, $\tau=0.5,\,\,1.0,\,\,2.0$ for panels from left to right, the height $h=1.0$ $r_g$, the inner and outer radii of the disc $R_{\rm in}=6.0$ $r_g$ and $R_{\rm out}=200.0$ $r_g$.}
\label{Fig11}
\end{figure*}

%~~~~~~~~~~~~~~~~~~~~~~~~~~~~~~~~~~~~~~~~~~~~~~~~~~~~~~~~~~~~~~~~~~~~~~~~~~~~~~~~~~~~~~~~~~~
%~~~~~~~~~~~~~~~~~~~~~~~~~~~~~~~~~~~~~~~~~~~~~~~~~~~~~~~~~~~~~~~~~~~~~~~~~~~~~~~~~~~~~~~~~~~

\begin{figure*}[t]
\center 
\includegraphics[scale = 0.5]{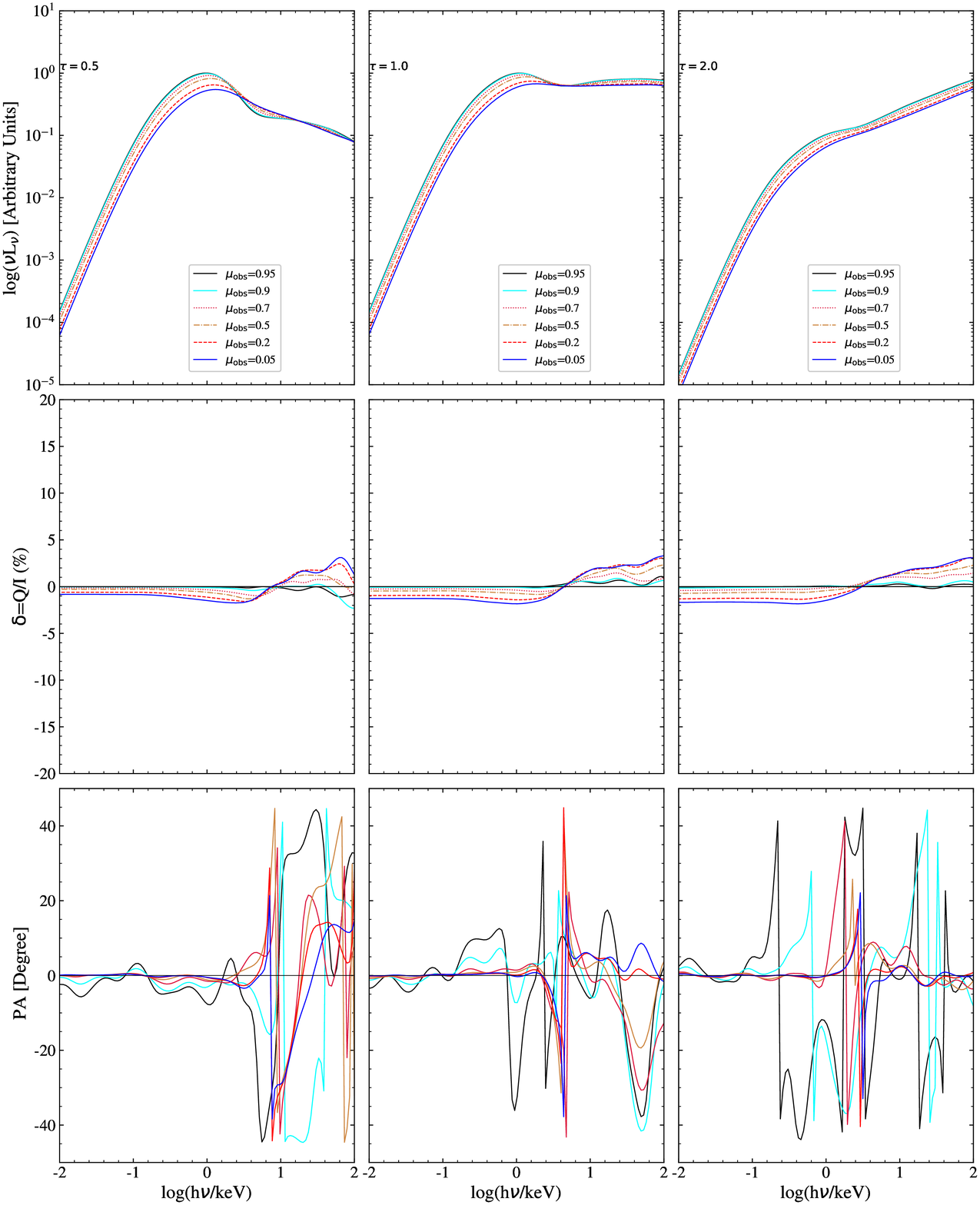} 
\caption{The same as Figure \ref{Fig11}, but for a corona with sphere geometry. The parameters for the sphere are: the radius of the sphere $R_{\rm sp}=200.0$ $r_g$, the height of the sphere center above the disc $h=0.0$ $r_g$.}
\label{Fig12}
\end{figure*}

%~~~~~~~~~~~~~~~~~~~~~~~~~~~~~~~~~~~~~~~~~~~~~~~~~~~~~~~~~~~~~~~~~~~~~~~~~~~~~~~~~~~~~~~~~~~
%~~~~~~~~~~~~~~~~~~~~~~~~~~~~~~~~~~~~~~~~~~~~~~~~~~~~~~~~~~~~~~~~~~~~~~~~~~~~~~~~~~~~~~~~~~~

\begin{figure*}[t]
\center 
\includegraphics[scale = 0.5]{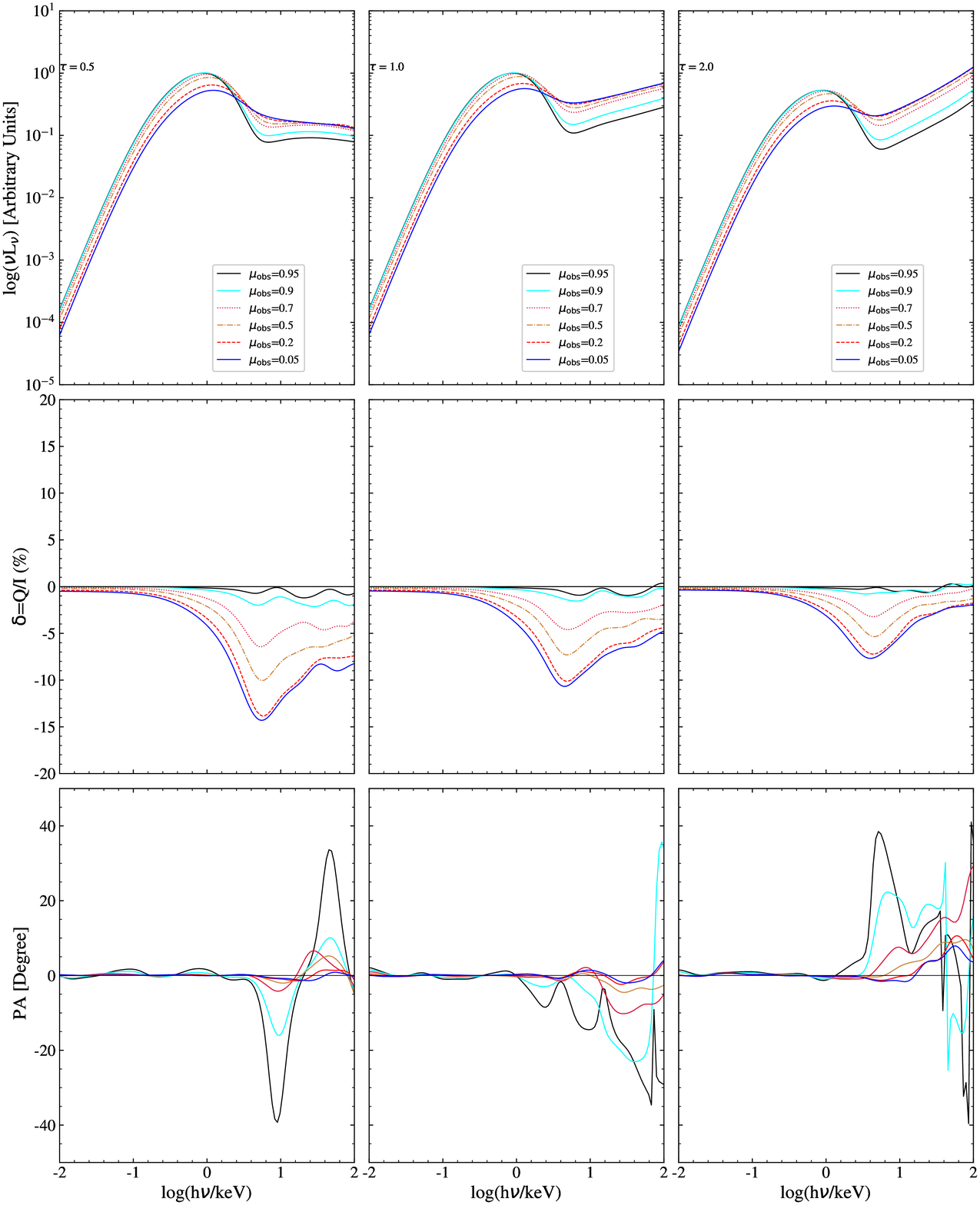} 
\caption{The same as Figure \ref{Fig11}, but for a corona with cylinder geometry. The parameters for the sphere are: the radius  $R_c=10$ $r_g$, the height $H=20$ $r_g$ and the intrinsic height  $H_c=100$ $r_g$.}
\label{Fig13}
\end{figure*}

\begin{figure*}[t]
\center 
\includegraphics[scale = 0.35]{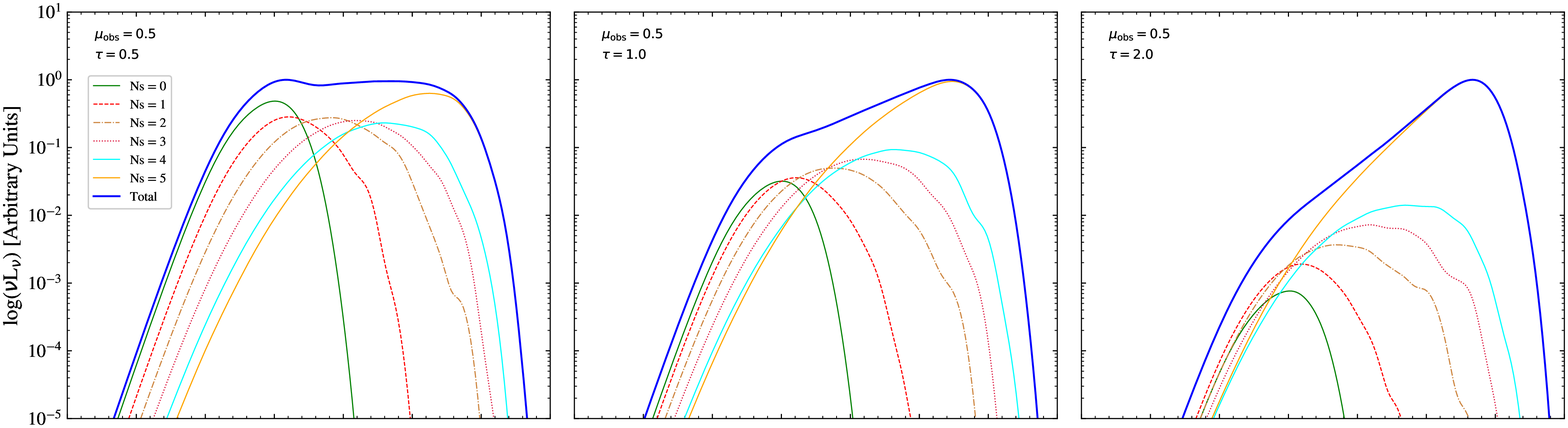}
\includegraphics[scale = 0.35]{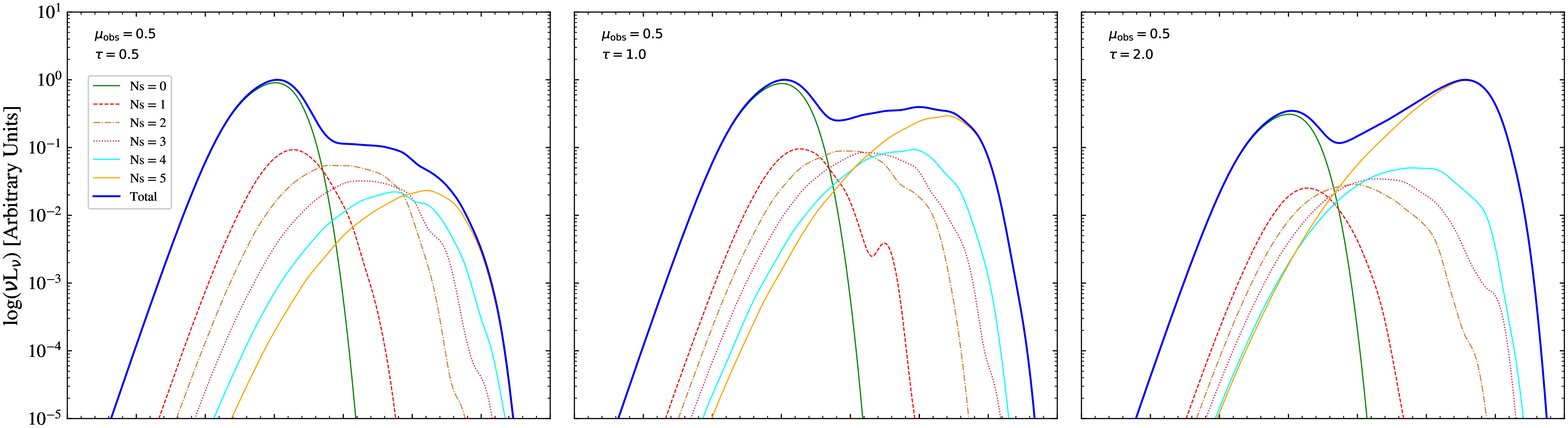} 
\includegraphics[scale = 0.35]{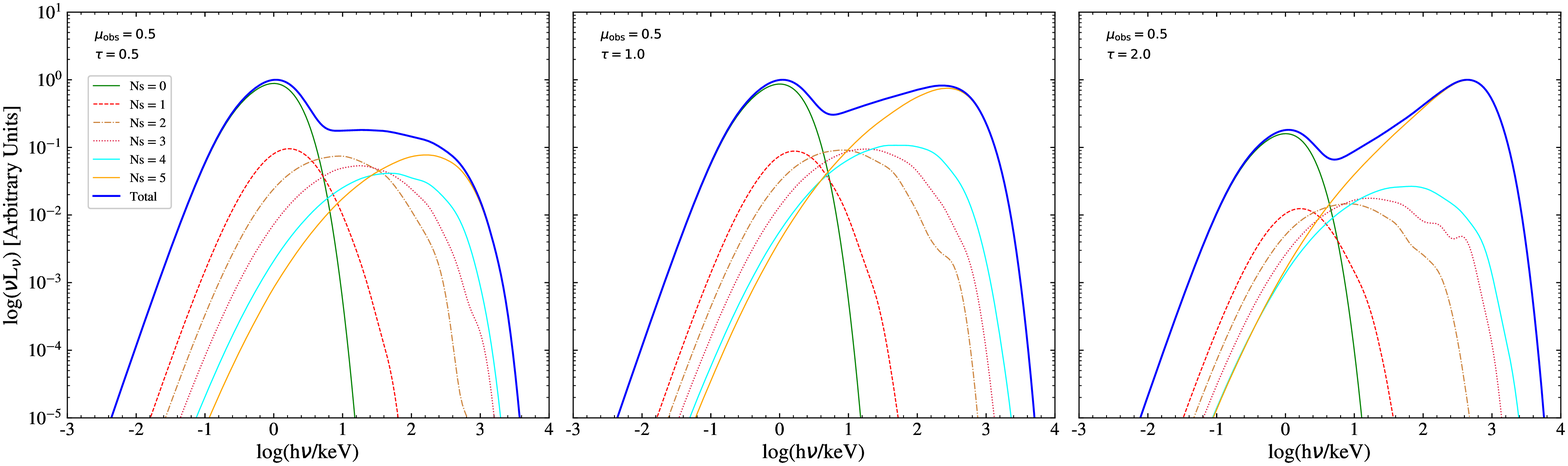}  
\caption{The synthesized spectra of coronas with a slab (top panel), sphere (middle panel) and cylinder (bottom panel) geometries for various scattering numbers. The cosine of viewing angle is  $\mu_{\rm obs}=0.5$. Other parameters are the same as that given in Figure \ref{Fig11}, \ref{Fig12} and \ref{Fig13} for the three kinds of coronas, respectively.}
\label{Fig14}
\end{figure*}

We first present the spectra, PDs and PAs of the radiations emerging from three kinds of coronas. The results are plotted in Figures \ref{Fig11}, \ref{Fig12} and \ref{Fig13}, which show that as the optical depth increases, the spectra become harder for all the cases. The profiles of the spectra are similar for three cases and also for different inclination angles. This supports the perspective of distinguishing the geometries for the coronas with different configurations through fitting the spectral energy distributions is not so effective (see also \cite{2018A...619A.105T, 2021ApJ...906...18D}). The polarizations of the three geometrical configurations, however, show significant differences both in the profile and the magnitude. One can see that in the X-ray bands, which we are mostly interested in, the slab corona has the biggest PD, whose value can be up to around 10 $\%$ depending on the viewing inclination. And followed are the magnitude of PD for the sphere and cylinder coronas, which goes to 1-2 \% and is less than 1 $\%$, respectively. These results are consistent with those given by \cite{2022MNRAS.510.3674U}. However, in the high energy bands, the results change into the opposite situation, where the cylinder corona has the highest PD up to almost 15 $\%$.  One can see that there exists zero points in PDs both for the slab and sphere coronas, as shown in Figures \ref{Fig11} and \ref{Fig12}, where the $Q$ component of the SPs vanishes and changes its sign simultaneously. From the parameters in Figures \ref{Fig11} and \ref{Fig12}, we can see that both of the two coronas have extending geometrical configurations. This may be taken as a special feature for this kind of coronas.

From Figures \ref{Fig11} and \ref{Fig12}, one can conclude that the trends and profiles of PD for the slab and the sphere are similar to each other but different to that of the cylinder case. This is because the former two coronas have similar geometrical configurations due to the parameters we chosen, i.e., they both have the extended configurations that fully cover the disc. On the contrary, the cylinder corona has a compact configuration above the disc, which makes its irradiation by disc to be less isotropical compared with the extending configurations. For the extending corona, a higher optical depth yields a higher PD, while for the compact corona, the conclusion seems opposite (see Figure \ref{Fig13}, where the maximum of PD decreases as optical depth increases).

According to our definition of the reference direction of the PA, for all the cases, the PA oscillates around zero and the polarization direction is horizontal in the low energy bands. This result is simply due to the fact that all of the geometrical configurations considered here are the axial-symmetry, which will yield an orientation of the polarization vectors that can either be horizontal or vertical \citep{1980ApJ...235..224C}. However, in the high energy bands, the PA oscillates turbulently to lead the results unreliable.

In Figure \ref{Fig14}, we give the spectra of all configurations for various scattering numbers and optical depths. For the spherical and cylindrical configurations, the profiles of the spectra are quite similar for different optical depth. But for the slab configuration, the spectra become harder as the optical depth increases.

\begin{figure*}[t]
\center 
\includegraphics[scale = 0.45]{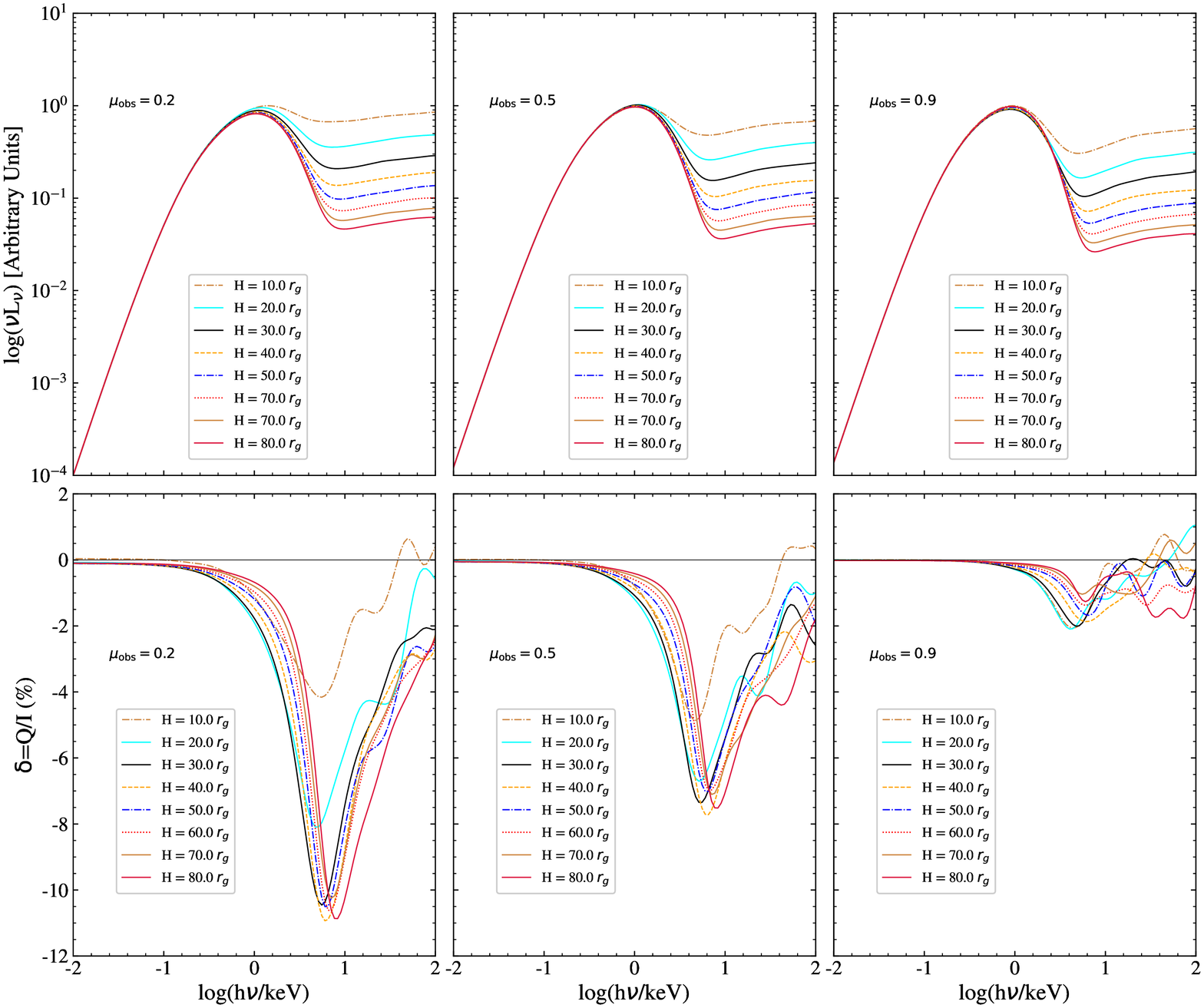} 
\caption{The spectra (top row panels) and PDs (bottom row panels) of a corona with a sphere geometry for different heights $H$ and viewing angles $\mu_{\rm obs}$. The parameters are: the radius $R_{sp}=10.0$ $r_g$, the potical depth $\tau=1.0$ and temperature $kT_e=100$ keV. The height $H$ of the spherical center varies from $10.0$ to $80.0$ $r_g$ and the corresponding results are plotted with different color and line styles. The cosine of the viewing angles $\mu_{\rm obs}$ is $0.2$, 0.5 and 0.9 for panels in the left, middle and right columns, respectively.}
\label{Fig21}
\end{figure*}

\begin{figure*}[t]
\center 
\includegraphics[scale = 0.48]{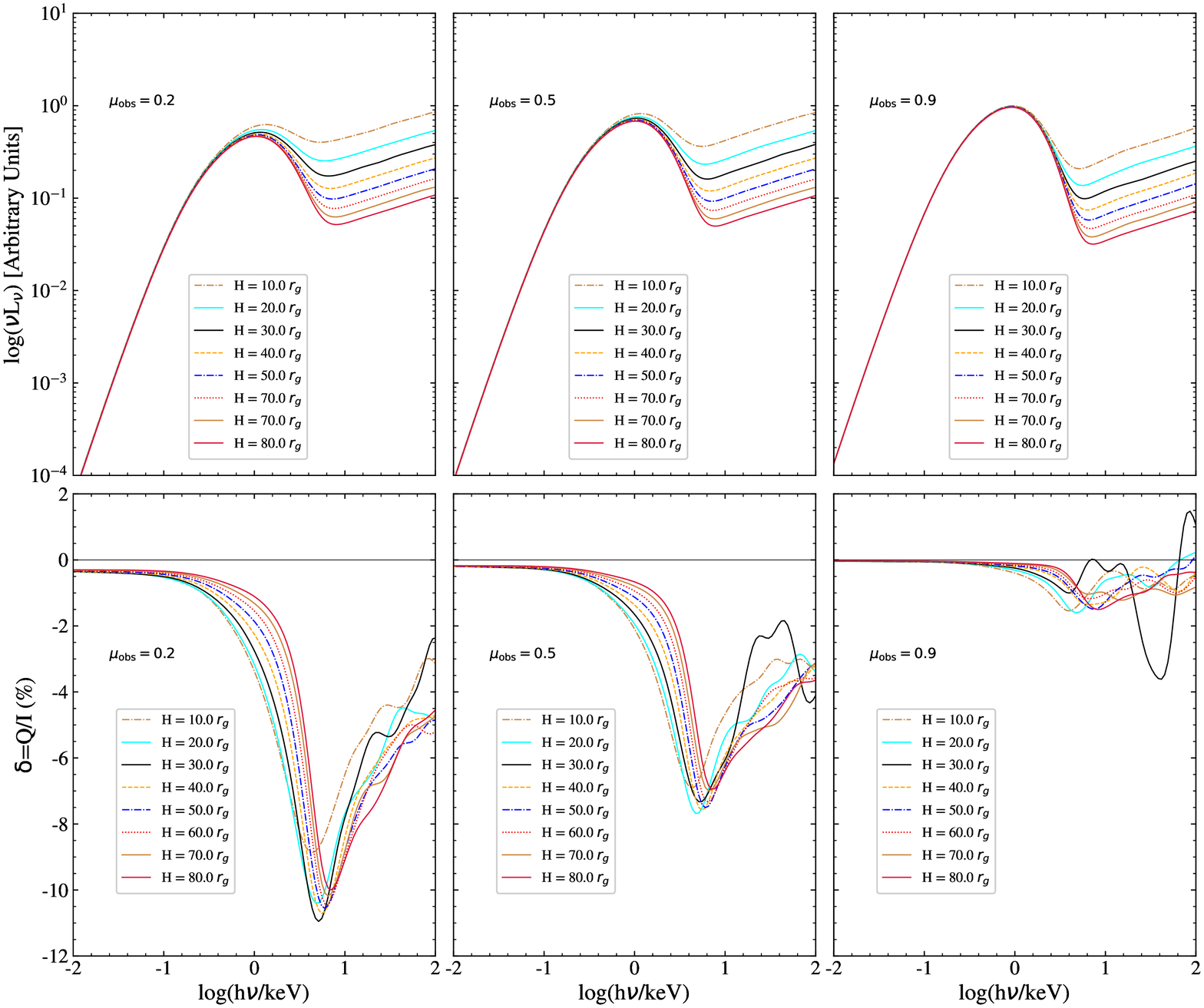} 
\caption{The same as Figure \ref{Fig21}, but for a corona with cylinder geometry. The parameters are: the radius  $R_c=10$ $r_g$ and the intrinsic height  $H_c=100$ $r_g$.}
\label{Fig22}
\end{figure*}

For the corona with a spherical or a cylindrical geometry, its height $H$ measured from the disc is an important parameter. The changes of $H$ will not only alter the geometry of the system, but also affect the efficiency that the disc illuminates the corona. For a higher $H$, the illumination is less isotropical. Hence we do simulations to investigate the impact of $H$ on the spectra and PDs. The results are shown in Figures \ref{Fig21} and \ref{Fig22} for the spherical and cylindrical cases, respectively. From these figures one can see that as $H$ increases, the spectra become softer in the high energy bands, due to the reduction of the luminosity and flux that could be captured by the corona. As the viewing angles change, the spectra tend to keep the same. But for the PDs, the magnitude of variations is considerable. Avaragely, the PDs of the cylinder corona are bigger than that of the sphere corona.
The profiles of the PDs for the two configurations are analogous, due to the quite similar geometrical parameters we choose. However, for the sphere case, with the increase of $H$, the $Q$ component will change its sign from positive to negative. But for the cylinder case, the sign of $Q$ always keeps negative. Also one can see that with the increase of $H$, the PDs in the low energy bands keep almost unchanged, just as the trend of the spectra. However, the changes of the PDs in the high energy bands are significant. This difference of PDs between the low and high energy bands may be used as a probe to discriminate a corona with or without a compact geometry.
%~~~~~~~~~~~~~~~~~~~~~~~~~~~~~~~~~~~~~~~~~~~~~~~~~~~~~~~~~~~~~~~~~~~~~~~~~~~~~~~~~~~~~~~~~~~
%~~~~~~~~~~~~~~~~~~~~~~~~~~~~~~~~~~~~~~~~~~~~~~~~~~~~~~~~~~~~~~~~~~~~~~~~~~~~~~~~~~~~~~~~~~~

\section{Discussion and Conclusions}
\label{Sec: discussions}

The geometrical configurations of corona in the accreting systems are poorly understood due to the degeneracy of spectroscopy differentiating the system. The polarimetry provides a more effective and powerful option to overcome this dilemma. With the advent of the missions of IXPE \citep{2016SPIE.9905E..17W}, eXTP \citep{zsn2016, zsn2019}, the era of high-quality data of polarization observations is arriving in the near future. Hence it is necessary and urgent to make the theoretical studies in advance.. For this purpose, we carry out the simulations of radiative transfer in corona with different geometries. These simulations are based on the public available Monte Carlo code, Lemon \citep{2021ApJS..254...29X}, which is based on the Neumann series expansion solution of differential-integral equations. By using the code, one can increase the signal to noise ratio dramatically and simplify the calculations when the configuration of the system has geometric symmetries. The main contents and results of this paper can concluded as follows:

We have discussed detailedly how to simulate the polarized radiative transfer in the three geometry configurations, namely, the slab, sphere and cylinder. We emphasized how to generate photons efficiently for the sphere and cylinder cases. To accomplish this, we have derived the shapes of the regions formed by the effective momentum directions in the $\phi$-$\mu$ plane of the triad. This method can increase the efficiency of the simulation, especially for coronas with a compact geometry, since the solid angle subtended by the corona with respect to the emission site is small, which further reduces the probability that a photon can be received by the corona. In our model, we considered the effect due to the Keplerian motion of the disc on the spectra. This effect works by affecting the photon generation. In our model, it can be taken into account readily through a Lorentz transformation, which connects the emissivities in the comoving reference frame of the disc and the static reference frame. The Keplerian motion of the disc will make the frequencies of the seed photons to have blue or red shifts, which will further affect the final results. While due to the low speed for most part of the disc ($R\gg 1$), the Keplerian motion seems to have a minor impact on the spectra and polarizations. However if we focus on the radiations emitted from the most inner part of the disc, both the Keplerian motion and general relativity effects should be taken into account necessarily. Then we discussed how to obtain the scattering distance between any two scattering sites by the inverse CDF method. Because the electron distribution has a very important impact on the Comptonization spectra, we proposed a new scheme to deal with three often used distribution functions, i.e., the thermal, $\kappa$ and power law, in a uniformed way. Next, we demonstrated how to implement the polarized Compton scattering in the Klein-Nishina regime consistently and detailedly, which involves the complicated triads constructions, Lorentz boosts, SPs transformation and rotations. Finally, we discussed the procedure for evaluating the contributions made by any scattering site to the observed quantities when the inclination angle of the observer is provided.

We use our model to simulate the radiative transfer in three kinds of coronas with different geometrical configurations. The results demonstrate that the polarizations of the observed radiations are significantly dependent on the geometries of the corona. Different configurations will produce PDs with different magnitudes and profiles in the X-ray bands. The corona with an extending configuration, such as the slab, yields a higher PD while the compact one yields a less polarized result. With the increase of the photon energy, the PD will increase gradually as well untill a maximum is reached. After that, the PD will decrease to zero due to the relativistic beaming effect \citep{2021ApJ...906...18D}. The maximum of PD for the extending configurations increases with the increase of the optical depth, but for the compacting configurations the conclusion is the opposite.
 
Our results are consistent with those of the former researches. However, our model and code are flexible and can deal with different geometrical configurations readily. One just needs to modify the photon generation and tracing parts. However, the results presented here are quite theoretical and not connected with the practical observations. Also, our model does not include the effects of the general relativity, which inevitably plays an important role in the radiative transfer around a black hole. These effects will be included in the future work. Nonetheless, our model includes the essetial ingredients of Comptonization in a hot electron corona. Thus, hopefully, our model will provide some useful insights for the observations of the upcoming X-ray missions, such as $XIPE$ and $eXTP$.

\section{Acknowledgments}
We thank the anonymous referee for helpful comments and suggestions that improve 
the manuscript significantly. We thank the Yunnan Observatories Supercomputing Platform, on
which our code was partly tested. We acknowledge the
ﬁnancial support from the National Natural Science Foundation
of China U2031111, 11573060, 12073069, 11661161010 and 11673060.

\section{code AVAILABILITY} 
The source code of this article is part of Lemon and can be downloaded from: \url{https://bitbucket.org/yangxiaolinsc/corona_geometry/src/main/}.

\label{lastpage}

\end{document}